\documentstyle[preprint,amstex,aps,eqsecnum,epsfig]{revtex}
\tighten
\def\JPA#1{{\sl J.\ Phys.\ {\bf A#1}}}
\def\IMPA#1{{\sl Int. J. Mod. Phys. {\bf A#1}}}
\def\IMPB#1{{\sl Int. J. Mod. Phys. {\bf B#1}}}
\def\NPB#1{{\sl Nucl.\ Phys.\ {\bf B#1}}}
\def\PLB#1{{\sl Phys.\ Lett.\ {\bf #1B}}}
\def\PRL#1{{\sl Phys.\ Rev.\ Lett.\ {\bf #1}}}
\def\ZPB#1{{\sl Z.\ Phys.\ B\ {\bf #1}}}
\def\CMP#1{{\sl Comm. Math. Phys. {\bf #1}}}

\def\a{\alpha}

\def\b{\beta}

\def\d{\delta}
\def\g{\gamma}

\def\e{\epsilon}
                       
\def\l{\lambda}

\def\s{\sigma}
\def\om{\omega}
\def\Om{\Omega}
\def\t{\theta}

\def\Th{\Theta}
\def\o{\over}

\def\lowmp{\lower.11em\hbox{${\scriptstyle\mp}$}}
\def\intf{\int_{-\infty}^{+\infty}}
\def\Im{{\rm Im\,}}
\def\Re{{\rm Re\,}}
\def\sign{{\rm sign}}

\def\der#1#2{{d#1\over d#2}}
\def\Q{{\cal Q}}

\def\zV{{z_{\mathrm V}}}
\def\zVpri{{z'_{\mathrm V}}}

\def\zH{{z_{\mathrm H}}}

\def\zHppri{{z'_{\mathrm H\,+}}}
\def\zHpm{{z_{\mathrm H\,\pm}}}

\def\zS{{z_{\mathrm S}}}

\def\zSppri{{z'_{\mathrm S}}}
\def\zSpm{{z_{\mathrm S\,\pm}}}

\def\zC{{z_{\mathrm C}}}
\def\zCpri{{z'_{\mathrm C}}}
\def\zCpm{{z_{\mathrm C\,\pm}}}

\def\zcppri{{z'_{\mathrm close\,+}}}

\def\zwppri{{z'_{\mathrm wide\,+}}}

\def\NH{{N_{\mathrm H}}}
\def\effNH{{N_{\mathrm H,eff}}}
\def\effNHpm{{N_{\mathrm H,eff}^\pm}}
\def\NHp{{N_{\mathrm H}^+}}

\def\NHm{{N_{\mathrm H}^-}}
\def\NHpm{{N_{\mathrm H}^\pm}}

\def\NHO{{N_{\mathrm H}^0}}

\def\NS{{N_{\mathrm S}}}
\def\NSp{{N_{\mathrm S}^+}}
\def\NSm{{N_{\mathrm S}}^-}
\def\NSpm{{N_{\mathrm S}^\pm}}
\def\NSO{{N_{\mathrm S}^0}}

\def\NB{{N_{\mathrm B}}}

\def\MR{{M_{\mathrm R}}}
\def\MC{{M_{\mathrm C}}}
\def\MRp{{M_{\mathrm R}^+}}

\def\MRm{{M_{\mathrm R}^-}}
\def\MRpm{{M_{\mathrm R}^\pm}}

\def\Mc{{M_{\mathrm close}}}
\def\Mcp{{M_{\mathrm close}^+}}

\def\Mcpm{{M_{\mathrm close}^\pm}}
\def\McO{{M_{\mathrm close}^0}}

\def\Mw{{M_{\mathrm wide}}}
\def\MwO{{M_{\mathrm wide}^0}}
\def\Mwp{{M_{\mathrm wide}^+}}
\def\Mwm{{M_{\mathrm wide}^-}}
\def\Mwpm{{M_{\mathrm wide}^\pm}}
\def\Mwup{{M_{\mathrm wide,\uparrow}}}
\def\Mwupm{{M_{\mathrm wide,\uparrow}^-}}
\def\Mwdown{{M_{\mathrm wide,\downarrow}}}
\def\Mwdownp{{M_{\mathrm wide,\downarrow}^+}}

\def\Mscpri{{M'_{\mathrm s-c}}}
\def\Mscpripm{{M_{\mathrm s-c}^{\prime\,\pm}}}

\def\WV{{W_{\mathrm V}}}
\def\WH{{W_{\mathrm H}}}
\def\WC{{W_{\mathrm C}}}

\def\IR{{I_{\mathrm R}}}
\def\IH{{I_{\mathrm H}}}

\def\IHp{{I_{\mathrm H}^+}}
\def\IHm{{I_{\mathrm H}^-}}
\def\IHpm{{I_{\mathrm H}^\pm}}

\def\IS{{I_{\mathrm S}}}
\def\ISp{{I_{\mathrm S}^+}}
\def\ISpm{{I_{\mathrm S}^\pm}}

\def\IC{{I_{\mathrm C}}}
\def\ICp{{I_{\mathrm C}^+}}
\def\ICpm{{I_{\mathrm C}^\pm}}

\def\Icp{{I_{\mathrm close}^+}}

\def\Iwp{{I_{\mathrm wide}^+}}

\def\Imax{{I_{\mathrm max}}}
\def\Imin{{I_{\mathrm min}}}
\def\Immax{{I_{\mathrm max}^-}}

\def\Ipmin{{I_{\mathrm min}^+}}

\def\DS{\triangle S}

\def\II{{\mathrm II}}

\def\EV{{E_{\mathrm V}}}
\def\EHp{{E_{\mathrm H}^+}}
\def\ESp{{E_{\mathrm S}^+}}
\def\ECp{{E_{\mathrm C}^+}}
\def\EKp{{E_{\mathrm K}^+}}
\def\EHpm{{E_{\mathrm H}^\pm}}
\def\ESpm{{E_{\mathrm S}^\pm}}
\def\ECpm{{E_{\mathrm C}^\pm}}
\def\EKpm{{E_{\mathrm K}^\pm}}

\begin{document}
\preprint{ LPTHE-96-46/; IFUMilano-546/FT \bigskip}
\draft
\title{\bf NON LINEAR INTEGRAL EQUATION AND
EXCITED--STATES SCALING FUNCTIONS  IN THE SINE-GORDON MODEL}
\author{{\bf C. Destri$^{(a)}$}  and {\bf H. J. de Vega$^{(b)}$\bigskip}}

\bigskip

\address
{ (a) Dipartimento di Fisica, Universit\`a di Milano
          and INFN, sezione di Milano
     \footnote{mail address: 
           Dipartimento di Fisica 
           Via Celoria 16, 20133 Milano ITALIA}\\
(b)  Laboratoire de Physique Th\'eorique et Hautes Energies, Paris
  \footnote{mail address: LPTHE, 
Universit\'e Paris VI, Tour 16, 1er \'etage, 4, Place Jussieu
75252 Paris, Cedex 05, FRANCE. Laboratoire Associ\'e au CNRS UA 280.\\}}

\date{January 1997}
\maketitle
\begin{abstract}
The NLIE (the non-linear integral equation equivalent to the Bethe
Ansatz equations for finite size) is generalized to excited states,
that is states with holes and complex roots over the antiferromagnetic
ground state. We consider the sine-Gordon/massive Thirring model
(sG/mT) in a periodic box of length $ L $ using the light-cone
approach, in which the sG/mT model is obtained as the continuum limit
of an inhomogeneous six vertex model. This NLIE is an useful starting
point to compute the spectrum of excited states both analytically in
the large $ L $ (perturbative) and small $ L $ (conformal) regimes as
well as numerically.

We derive the conformal weights of the Bethe states with holes and
non-string complex roots (close and wide roots) in the UV limit.
These weights agree with the Coulomb gas description, yielding a
UV conformal spectrum related by duality to the IR conformal spectrum
of the six vertex model.

\end{abstract}

\section{Introduction}
 
The NLIE proposed in ref.\cite{prl} allows to treat in an unified way
the thermodynamics of magnetic chains and the finite size corrections
to vertex models solvable by Bethe Ansatz.  Moreover, in
ref. \cite{prl} we derived (using the light-cone approach) the NLIE
that describes the ground state of the sine--Gordon/massive--Thirring
(sG/mT) field theory on a finite spatial volume (with periodic
boundary conditions). The NLIE has been successfully used in different
situations \cite{usoddv,blz,npb}.

In the present paper we generalize the NLIE to {\em excited
states}. That is, we derive the nonlinear integral equation equivalent
to the Bethe Ansatz equations for states with holes and complex BA roots
around the antiferromagnetic ground state in a periodic box of size
$L$ (the case with only holes in certain special configurations has
been treated also in \cite{rava}). 
In our framework the complex roots do not generally appear in the form of
Bethe or Takahashi strings \cite{hlba}. That is, their imaginary parts take
continuous values (even for infinite volume) which are determined by the BAE 
themselves.

One can derive in an analogous way the NLIE for the finite--size
effect on {\em excited states} in the six vertex model, as done in
\cite{klum} and the so--called excited--states thermodynamics of the
XXZ chain at temperature $T=1/L$.

NLIE closely related to ours are obtained in ref. \cite{blz} along
different lines which starts from the Perturbed Conformal Field
Theory in the continuum. This method, however, is not yet 
directly applicable to the sG/mT model.

As is known \cite{clmtm,rev}, the sine--Gordon model with 
coupling $\beta$ admits an integrable U(1)--invariant light--cone
lattice regularization based on the $R-$matrix of the six--vertex
model with anisotropy $\g=\pi-\beta^2/8$. The energy--momentum
spectrum is then calculated exactly by means of the Algebraic Bethe
Ansatz, or Quantum Inverse Scattering Method: a Bethe Ansatz state is
identified by an unordered set of distinct, generally complex numbers
$\l_1,\l_2,\ldots,\l_M$ which satisfy the famous Bethe Ansatz equations
\begin{equation}\label{eq:baeTh}
	\left[ {{s_\g(\l_j + \Th + i\pi/2)} \o 
		{s_\g(\l_j + \Th - i\pi/2)}}\right]^N   
	\left[ {{s_\g(\l_j - \Th + i\pi/2)} \o
		{s_\g(\l_j - \Th - i\pi/2)}}\right]^N =
	-\prod_{n=1}^M {{s_\g(\l_j -\l_n +i\pi)}
		 \o {s_\g(\l_j-\l_n -i\pi)}} 
\end{equation}
where $ s_{\g}(x)=\sinh(\g x/\pi), \;  N $ stands for the number of sites, 
$ L= N \delta $ is the physical size of the system (with
periodic boundary conditions) and $\delta$ is the lattice spacing (that
is the inverse of the UV cutoff). The energy $E$ and momentum $P$ of
this BA state can be extracted from the relation
\begin{equation}\label{eq:expEP}
      e^{-i(E\pm P)\delta/2}= \prod_{j=1}^M
	{{s_\g(i\pi/2+\Th \pm \l_j)}\o{s_\g(i\pi/2-\Th \mp \l_j)}} 
\end{equation} 
The real parameter $\Th$ plays the role of rapidity cutoff and will
diverge in the continuum limit $ \d\to 0 $ in such a way to keep the the
physical mass scale $m$ fixed. 

To be precise, the continuum relativistic QFT defined on the infinite
Minkowski plane follows by first taking the IR limit $L\to\infty$
(that is $N\to\infty$ at fixed $\delta$) and then the UV limit
$\delta\to 0$ near the critical point $\Th=\infty$, holding $m \sim
\d^{-1}\exp(-\Th)$ fixed. On the other hand, by taking the
continuum limit ($ \d\to 0 $) at fixed $L$, we get instead the same
QFT on a ring of length $L$. It is this second procedure that we wish
to study here. 

Let us also recall that the sG model has two distinct regimes, one
repulsive, for $0<\g<\pi/2$, and one attractive, for $\pi/2<\g<\pi$.
In the repulsive regime the spectrum contains only solitons and
antisolitons, with U(1) charge $ S = +1/2$ and $-1/2$ respectively (this
charge is properly quantized w.r.t. the nonlocal hidden SU(2)$_q$
symmetry of the model \cite{resh}), while in the attractive regime there are
also neutral bound state of these, the so-called breathers. In the BA
solution the soliton/antisoliton states appears as holes in the ground
state distribution of BA roots; the breather states appear instead as
special configurations of complex roots (see below). 

We  treat in an unified way both the repulsive and the attractive regimes.

The central object in the NLIE is the counting function $ Z(\l) $.
Its name follows from the fact that $ \frac1{2\pi} \; Z(\l_j) $ is an
integer for odd $ S $ and a half-odd integer for even $ S $ at the
roots $ \l_j $ of the BAE.  In addition, $ Z(\l) $ is monotonically
increasing in the bulk.  $ Z(\l) + \frac14(1 + (-1)^S ) $ can take
values which are integer multiple of $2\pi$ also for real $ \l $ which
are not roots of the BAE. These are the so called holes and together
with the complex roots describe the excited states.

We find that  $ Z(\l) $ may be {\em decreasing} at some roots and holes.
We call such points {\em special} roots/holes. They appear in the 
borders of the bulk where the root density becomes sparse.
The presence of special real roots/holes turns to be a crucial feature in the
analysis of the excited states.

The NLIE takes the following form in the sG model for arbitrary
excited states  with U(1) charge $S$ (notice that the adopted periodic
boundary conditions force $S$ to be an integer)
\begin{equation}\label{eq:ddv}
	Z(\l) = mL\sinh\l + g(\l) 
		+ \intf dx\, G(\l -x)\,\Q(x) \; ,
\end{equation}
where the unknown $Z(\l)$ is the counting function,
$\Q(x)$ is the nonlinearity
\begin{equation*}
	\Q(x) =	-i\,\log
	\dfrac{1+ (-)^S\,e^{i Z(x+i\e)}}{(-)^S+ e^{-i Z(x-i\e)}} 
\end{equation*} 
and the function $g(\l)$ contains the information about the excited
state considered
\begin{equation}\label{eq:gintro}
	g(\l)= \zH(\l) + \zS(\l) + \zC(\l) \; ,
\end{equation}
where $ \zH(\l), \;   \zS(\l) $ and $ \zC(\l) $ stand for the contribution
of the holes, the special holes  and the complex
roots. 

We have,
\begin{eqnarray}
	\zH(\l) &=&\sum_{j=1}^\NH \chi(\l-h_j) \cr \cr
\zS(\l) &=& -2\sum_{j=1}^\NS \chi(\l-y_j) 
\end{eqnarray}
where the $ h_j $ stand for the positions of the holes and $ y_j $ for
those of the special root/holes.  The form of $ \zC(\l) $ depends
whether $ \g < \pi /2 $ (repulsive regime) or $ \g >\pi /2 $
(attractive regime).  It is given in eqs. \eqref{eq:zCR} -
\eqref{eq:zCA}.

The kernel $ G(\l-x) $, explicitly written in eqs. \eqref{eq:G}, is just
$(2\pi)^{-1}$ times the logarithmic derivative of 
the soliton--soliton scattering amplitude.

For the ground state, the NLIE  \eqref{eq:ddv} reduces to the form
presented in \cite{prl,npb}.

A new way to write the NLIE follows by explicitly performing the
limit $\e\to 0$ in eq. \eqref{eq:Q}:
\begin{equation}\label{eq:Qmod}
	\Q(x) = \{Z(x) +\tfrac12[1-(-1)^S]\pi \} \text{ mod } 2\pi    
\end{equation}
where the mod $2\pi$ restriction may  be written as
\begin{equation}\label{eq:defmod}
	X \text{ mod } 2\pi =  X - 2\pi \,\sign(X)
	\left\lfloor \dfrac{|X|}{2\pi}+\dfrac12 \right\rfloor  
\end{equation}
for any real number $X$ ($\lfloor x \rfloor$ stands for the integer
part of $x$).  

Using eqs.\eqref{eq:Qmod} and \eqref{eq:defmod} reduces the nonlinearity
in eq.\eqref{eq:ddv} to an integer part calculation. We use this new form
of the NLIE to solve it for $mL \to 0$ and $ \Th = \pm \infty $ in sec. VII.
Namely, in the plateau regions where the counting function is flat the NLIE
reduces to a simple algebraic equation. For instance, as $ \l \to \infty $ with
finite $ \Th $ we get [see eq. \eqref{eq:infNLIE}],
 \begin{equation}\label{eq:InfNLIE}
	X = b + \dfrac{\chi_\infty}\pi ( X \text{ mod } 2\pi )
\end{equation}
Where $X=Z_N(+\infty)+\delta_S \; \pi$, $ b $ is a known constant and 
$ \chi_\infty = {{\pi/2-\g}\o{1-\g/\pi}} $. Since $ X \text{ mod } 2\pi
 = X -2\pi n $ for a suitable integer $n$, 
 eq.\eqref{eq:InfNLIE} is solved immediately by
\begin{equation*}
	X = 2(1-\g/\pi)b - 2n(\pi-2\g)
\end{equation*}
 provided
$$
	|b - 2 \pi n | \le \dfrac\pi{2(1-\g/\pi)} \; .
$$

The energy--momentum for an arbitrary excited state can be expressed in terms
of the counting function as follows,
\begin{equation}\label{eq:EPcontI}
	E \pm P = \EV +  m\sum_{j=1}^\NH e^{\pm h_j} 
	 -2 \, m\sum_{j=1}^\NS e^{\pm y_j} + \ECpm
	\mp  m \intf {{dx}\o{2\pi}}\,e^{\pm x}\,\Q(x) 
\end{equation}
where $\EV$ is the ground state bulk energy,
\begin{equation*}
	 \EV = N\d^{-1} \left[-2\pi+\intf d\l\,
          {{\phi_{1/2}({{\pi \l}\o {\g}}+2\Th)}\o{\pi\; \cosh\l }}\right] 
\end{equation*}
 $ \EHpm $  and $ \ESpm $ stand for the contributions from holes and special
holes, respectively.
$\ECpm$ represents instead the contributions of the complex
roots. Its form depends whether  $ \g < \pi /2 $ (repulsive regime)
or  $ \g >\pi /2 $ (attractive regime). $\ECpm$ is given in 
 eqs.\eqref{eq:ContrC} and \eqref{eq:ECatt}, respectively.

\bigskip

We provide in the present paper a thorough analysis of the excited
states in the six vertex/sine-Gordon model as a basic step to derive
the NLIE.  We analyze the distribution of real roots for generic
excited states in secs. II and III for large but finite $ \Th $ and $
N $.  There are two bulk seas of real roots centered around $ \l = \pm
\Th $ and possibly a few isolated roots at the two extremities and in
between the two seas.  In such regions the counting
function exhibits a, possibly non--monotonic behaviour characterized
by one or two plateaus.

We derive a general relation for the spin 
in terms of the number of holes and complex roots [eq.\eqref{eq:NHetcI}].
In the continuum limit (sG model), it takes the form
$$
2 S = \NH -2\NS -\Mc - 2\,\Mw\; \theta(\pi-2\g)
$$
where $ \NH $ is the number of holes, $ \NS $ is the number of special holes,
$ \Mc $ the number of close roots and $ \Mw $ the number of wide roots. 
Notice that $ \NH -2\NS $ acts as the effective hole number.

Concerning the complex roots, our analysis fully uses the non-string
complex roots \cite{hlba}. Bethe strings do not appear as generic
solutions in our approach. These non-string complex roots do not
contribute to the energy--momentum in the repulsive regime. They have
different properties depending on their distance to the real axis
(close or wide roots).  They are just internal quantum numbers
describing the collective $ U(1) $ spin state of the excitation.  In
addition, wide roots in the attractive regime appear as independent
excitations not carrying any spin.  In the attractive regime, regular
arrays of complex roots that do contribute to the energy--momentum
appear in the infinite volume limit. Such arrays only contain wide
roots and describe hole bound states (breathers).

\bigskip

The NLIE  \eqref{eq:ddv} is therefore an useful starting point to find the 
spectrum of excited states both numerically or analytically in the large 
$ m L $ (perturbative) and small $ m L $ (conformal) regimes.

\bigskip

In section VIII we derive the scaling limit of the NLIE describing the 
conformal limit (UV regime) of the sG theory. This NLIE is simpler than the 
full sG-NLIE but it cannot be solved at present in closed form.
However, we succeed to compute the eigenvalues exactly in the conformal
regime for all excited states with the help of the Lemma 
\eqref{eq:lem}. This Lemma  yields the integral involved in the eigenvalue 
calculation in close form without knowing the solution of the
NLIE. We then show  that the sG model exhibits in the UV limit,
the conformal spectrum of a Coulomb gas with unit central charge 
and compactification radius
$$
	R= \sqrt{2  (1-\g/\pi)} = {{\b} \o {\sqrt{4\pi}}}
$$
The Bethe states provide the primary as well as the secondary conformal 
states. We find that the (UV) sG conformal spectrum is related with the
(IR) six-vertex conformal spectrum (which coincides with the  IR conformal
spectrum for the XXZ chain) 
by a duality transformation $ R \leftrightarrow R^{-1} $. 

Some simple excited states are discussed in detail in sec. VIII.

\section{Generalities}\label{general}

The BA roots are either real or come in complex conjugated pairs
$(\xi,{\bar\xi})$, with the exception of single roots at 
$\Im\l={{\pi^2} \over {2 \g }} $ or 
$\Im\l=-{{\pi^2} \over {2 \g }} $ , which are actually 
self--conjugated due to $ i {{\pi^2} \over { \g }}$
periodicity [see eqs. \eqref{eq:baeTh}-\eqref{eq:expEP}].
It is well known that the ground state, or vacuum in the
QFT language, corresponds to the unique BA solution with $N$ real
roots for the entire range $0<\g<\pi$. Then in any physically relevant
transfer matrix eigenstate there are $\MR$ real roots
$r_1,r_2,\ldots,r_\MR$ and 
$\MC=M-\MR$ complex roots $\xi_1,\xi_2,\ldots,\xi_\MC$ with $\MR$ of
order $N$ and $\MC$ of order one. Moreover, we recall that $S=N-M$ is
the eigenvalue of conserved U(1) charge (the third component of the
total spin in the six--vertex language).

Taking the logarithm of eq.\eqref{eq:baeTh} one finds,
\begin{equation}\label{eq:lnbae}
	N[\phi_{1/2}(\l_k+\Th)+\phi_{1/2}(\l_k-\Th)]
	-\sum_{j=1}^M \phi_1(\l_k-\l_j) = 2 \pi I_k 
\end{equation}
where the $I_k$ are half-odd-integers for $M$ even and integers for
$M$ odd and 
\begin{equation}\label{eq:defi}
	\phi_\nu(\l) \equiv
		i\log{{s_\g(i\nu\pi+\l)}\o{s_\g(i\nu\pi-\l)}}
	 =i\log{{\sinh[\g(i\nu+\l/\pi)]}\o{\sinh[\g(i\nu-\l/\pi)]}} 
\end{equation}
We choose the logarithmic cuts to run parallel to the real axis and
such that $ \phi_\nu(\l) $ is an odd function of complex $ \l $ (see
figs. 1 and 2). This removes any $2\pi$ ambiguity on the ``quantum
numbers'' $I_k$.  Notice that $ \phi_\nu(\l) $ is monotonic function.
Moreover, since $N$ may be chosen to be even (eventually
$N\to\infty$), the parity of $M$ is the same of $S+N-M$, so that the
$I_k$ are half-odd-integers for $S$ even and integers for $S$ odd.
Therefore we may write in general
\begin{equation*}
	I_k = \text{ integer } + \tfrac12 (1-\delta_S) \;,\quad 
	\delta_S \equiv \tfrac12[1-(-1)^S]
\end{equation*}

\begin{flushleft}
$\bullet$ {\bf The counting function}
\end{flushleft}

Given a solution $\l_1,\l_2,\ldots,\l_M$ of the BAE, we define the
corresponding {\em counting function} as 
\begin{equation}\label{eq:Zagain}
	Z_N(\l)=  N[\phi_{1/2}(\l+\Th)+\phi_{1/2}(\l-\Th)]
		   -\sum_{k=1}^M \phi_1(\l-\l_k)  
\end{equation}
Comparing eq. \eqref{eq:Zagain} to eq. \eqref{eq:baeTh} we have by
definition
\begin{equation}\label{eq:quant}
	Z_N(\l_k)= 2\pi I_k \;,\quad k=1,2,\ldots,M
\end{equation}
As stated above, the BA root $\l_1,\l_2,\ldots,\l_M$ must all be
distinct (the corresponding BA state would otherwise vanish). 
Once specialized to the real roots, eq. \eqref{eq:quant} reads
\begin{equation}\label{eq:quantr}
	Z_N(r_j)= 2\pi \IR_{\,j} \;,\quad j=1,2,\ldots, \MR	
\end{equation}
We naturally assume the ordering $r_1<r_2<\ldots<r_\MR$.
Furthermore, we anticipate that the size of the distribution of real roots
is of order $2\log N$ for large $ N $, 
that is $r_1\sim -\log N$, $r_\MR\sim +\log N$.

\begin{flushleft}
$\bullet$ {\bf Holes}
\end{flushleft}

Distinct real numbers $h_j$, $j=1,2,\ldots,\NH$, that are also
distinct from the real BA roots $r_1,r_2,\ldots,r_\MR$ but satisfy the
same quantization rule \eqref{eq:quantr}, that is
\begin{equation}\label{eq:quanth}
	Z_N(h_j)= 2\pi \IH_{\,j} 
\end{equation}
with the $\IH_{\,j}$ integers or half-integers, are called {\em holes}.
Again we assume the ordering $h_1<h_2<\ldots<h_\NH$.
Together with the real BA roots the holes form the complete set of
real zeroes of the function 
$$
1+(-1)^S e^{iZ_N(\l)} \; .
$$ 
We denote this set
as $\{x_j;\,j=1,2,\ldots,\MR+\NH\}$ and assume it ordered.

\begin{flushleft}
$\bullet$ {\bf Special root/holes}
\end{flushleft}

In the counting function \eqref{eq:Zagain} the term
$N[\phi(\l+\Th,\g/2)+\phi(\l-\Th,\g/2)]$, which is monotonically
increasing over the real line, acts as source of possible quantum
numbers for real roots and holes (a sort of phase space). Each summand
in the sum over roots either subtracts or add phase space depending on
the sign of $\pi-2\g$ and the imaginary part of the root. In general,
in any zero--temperature physical state, the global effect is to
produce a $Z_N(x)$ which is monotonically increasing on the real line,
but there could be exceptions, as we shall now discuss.

Let us call ``bulk regions'' the portions of the real line where
$Z_N'(x)$ is positive and of order $N$ for large $N$. In such regions
the spacing between consecutive real roots is of order
$1/N$. Therefore, the number of roots within the bulk is of order $N$.
The regions around the two extremities of the real distribution are by
definition not in the bulk and there $Z'_N$ may indeed change sign due
to one or more isolated roots. Another place where $Z'_N$ might change
sign is the middle of the real distribution, which is peaked around
$+\Th$ and $-\Th$. The limit $N\to\infty$ forces $Z_N(x)$ to be
monotonic increasing for any fixed $x$ and $\Th$. But for any
arbitrarily large $N$ there could be specific configurations of
quantum numbers and large enough values of $|x|$ and/or $\Th$, such
that $Z_N$ is decreasing at the two extremities and/or in the middle
of the real distribution. This possibilities are easily verified
numerically.

If $Z_N(x)$ decreases, it may decrease enough to cross
downward a quantization point already crossed upwards. Therefore the
following classification become necessary. We shall define as
{\em non-degenerate} those configurations of quantum numbers such that
the union
\begin{equation*}
	\IR \cup \IH = \{ \IR_{\,j};\,j=1,2,\ldots \MR \} \cup 
	\{ \IH_{\,j};\,j=1,2,\ldots \NH \}
\end{equation*}
contains only {\bf distinct} integers or half-odd-integers. This is
automatically the case if $Z_N'(x_j)>0$ at all zeroes of 
$$ 
1+(-1)^S e^{iZ_N(\l)}\; ,
$$ 
so that the ordering of the $ x_j$'s implies the
ordering of the corresponding quantum numbers $ I_j $. The {\em degenerate}
configurations are those where one or more quantum numbers appear more
than once in $\IR \cup \IH$. Evidently to such quantum numbers are
associated at the same time real zeroes of 
$$
1+(-1)^S e^{iZ_N(\l)}
$$
with $Z_N'(x_j)>0$ and real zeroes with $Z_N'(x_j)<0$. We call the
latter {\bf special} real roots or holes, root/holes for short, as
opposed to the normal ones with $ Z_N'(x_j)>0 $.  Notice that two
(or more) roots cannot be associated to the same integer. If this
would be the case, by continuity in $\g$ and $\Theta$, one could cause
these two roots to ``merge'' and thus obtain a double root of the BAE
which is to be discarded. Hence to a degenerate quantum number is
associated at most one real root, while the other are holes. The same
continuity argument in the two free parameters $ \Th $ and $ \g $ serves
to deal with the exceptional cases $ Z_N'(x_j)=0 $. We may regard such
cases as mergings of a real roots and a hole or of two holes. In
either case they do not require a special treatment.

Extensive numerical studies have shown that degenerate configurations
are restricted to few types. For large $N$ at fixed $\Th$ special
root/holes may appear only at either one of the two tails of the real
distribution, due to a single real or self--conjugated root or to a
cluster of complex roots that become isolated from the bulk when $\g$
gets close to one of a special set of rational multiples of $\pi$.  If
$\Th$ is large enough at fixed $N$, then the distribution of real
roots separates into two distinct distributions peaked around $+\Th$
and $-\Th$, respectively. In this case there could be special root/holes 
whenever one or more roots
remain at a distance of order $1$, rather than $\Th$, from the origin
and therefore become isolated. 

In this section we consider large $N$ at fixed $\Th$
and shall deal with the opposite regime of large $\Th$ at fixed $N$ in
section \ref{largeTh}.

Let us consider here an explicit example of root/holes in the outside
tails. We consider the BA states obtained by removing $S$ real roots
from the ground state distribution. If $2\g S<\pi$, then one finds
that there are $ N + S $ allowed values for the quantum numbers.
This follows from the limiting values
$$
\frac1{2\pi} Z_N(\pm\infty)=\pm\frac1{2\pi}[N\pi+(\pi-2\g)S] =
\pm [ {{N - S + 2 S}\over{2}} - \frac{\g S}{\pi} ]\; .
$$
 This is
just right to accommodate $N-S$ real roots (as required) and $2S$
holes. We are thus dealing with a non-degenerate configuration and
this is confirmed numerically (it can be done, for $N$ in the
thousands and to high precision, on any modern personal computer). One
finds also the following: if the smallest and largest of the $N+S$
real zeroes of
 $$
1+(-1)^S e^{iZ_N(\l)}
$$
 (say $x_1$ and $x_{N+S}$) are
holes, then $Z_N(x)$ is globally monotonic; if instead $x_1$ is a real
root, then $Z_N(x)$ tends to $Z_N(-\infty)$ from below and $Z_N'(x)$
changes sign just before $x_1$; if $x_{N+S}$ is a real root, then
$Z_N(x)$ tends to $Z_N(+\infty)$ from above and $Z_N'(x)$ changes sign
just after $x_{N+S}$. Suppose now we let $2\g S\to\pi^-$ while
keeping all quantum numbers fixed. One finds that when $x_{N+S}$ is
a hole, it simply goes to $+\infty$ and then ``jumps'' to the line
$\Im\l=-\pi^2/(2\g)$ (so that it ceases to be a real zero) when $2\g
S$ exceeds $\pi$; in the meantime $Z_N(x)$ remains monotonic in the
right tail. The same scenario applies to the left tail if $x_1$ is a
hole when $2\g S<\pi$, with the only difference that $x_1$ jumps to
the line $\Im\l=+\pi^2/(2\g)$. [The choice of sign in the two jumps is
a consequence of our logarithmic branch conventions for the function
$\phi_\nu(\l)$ and the request that all quantum numbers stay
fixed]. On the other hand, if $x_{N+S}$ is a root, then
$Z_N(+\infty)$ tends to $2\pi\IR_{\,N+S}$ as $2\g S\to\pi^-$, so
that we may say that a special hole appears at $+\infty$. When $2\g S$
exceeds $\pi$ this special hole moves in to a finite values $h$ until
it collides from the right with the root $x_{N+S}$ for a certain,
configuration--dependent value of $\g$. Then the root and the hole
split again, but now with $h<x_{N+S}$ and $Z_N'(h)<0$, while
$x_{N+S}$ is now a special root. 

Finally, as $\g$ approaches the
universal value $\pi/(S+1)$, $x_{N+S}$ tends to $+\infty$ and then
jumps to $\Im\l=-\pi^2/(2\g)$ when $\g$ exceeds $\pi/(S+1)$.  Notice
that the interval $\pi/(2S)\le\g\le\pi/(S+1)$ of non--monotonicity
shrinks to the single free--field point $\pi/2$ for $S=1$.  One can
also check that, after the jump, $h$ is still the largest real zero of
$Z_N(\l)$ when $S\ge 2$, while a new special hole appears when
$S=1$. A specular description applies in the left tail.

It should also be clear that the mechanisms just described may repeat
for larger values of gamma, provided we start from suitable
configurations. Likewise, it is possible that clusters of $n>1$
complex roots become isolated and are then ``pushed to infinity'' for
certain configurations and special values of $\g$. Looking at the
original BAE one sees that these complex roots must tend
asymptotically to form $q-$strings with the same real parts and
spacing $\pi^2/(q\g)$ in the imaginary direction \cite{rsos}.

\begin{flushleft}
$\bullet$ {\bf Complex roots}
\end{flushleft}

Besides the classification of real roots and holes into normal and
special types, it proves convenient to classify the
complex roots $\xi_j$, $j=1,2,\ldots \MC$ into {\em close roots}
\begin{equation*}
	\{c_j\,;\,j=1,2,\ldots,\Mc\} \;, \quad
	|\Im c_j| <\min(\pi,\pi(\tfrac{\pi}{\g}-1) )
\end{equation*}
and {\em wide roots}\cite{hlba}
\begin{equation*}
	\{w_j\,;\,j=1,2,\ldots,\Mw \} \;, \quad
	\min(\pi,\pi(\tfrac{\pi}{\g}-1)) <|\Im w_j| \le \tfrac{\pi^2}{2\g}
\end{equation*}
The  self--conjugate 
roots need to be further divided into two distinct classes. The first
class is formed by the real roots that have jumped to
$\Im\l=\pm{{\pi^2}\o {2 \g}}$ by passing through $\Re\l=\mp\infty$
upon suitably varying $\g$ at fixed quantum numbers, as we have
described in the example above. The second class is formed by all
other self--conjugated roots.  That is, the first class are those
 self--conjugated roots that become real roots if we make $ \g $ 
small enough keeping the state fixed. 

When $\g<\pi/2$ the total number, locations and quantum numbers of the
self--conjugated roots of the second class are connected to those of
the holes.  When $\g>\pi/2$ they are independent variables (see
eq. \eqref{eq:NHetc} below) and describe the lightest breather
states. In either case the presence of such roots properly
characterizes the BA state as an excited state and cannot be
eliminated just by varying $\g$.

\begin{flushleft}
$\bullet$ {\bf Relation among the various numbers of roots and holes }
\end{flushleft}

It is fairly easy to establish a relation among the U(1) charge and
the number of holes, special root/holes and complex roots.  From the
definition itself of the counting function, eq.\eqref{eq:Zagain} 
and the asymptotic values of the function $ \phi_\nu(\l) $ we read
\begin{equation}\label{eq:asym}
\begin{split}
	Z_N(+\infty) &= +N\pi + (\pi-2\g)S + 2\pi\sign(\pi-2\g)\Mwdown \\
	Z_N(-\infty) &= -N\pi - (\pi-2\g)S - 2\pi\sign(\pi-2\g)\Mwup 
\end{split}
\end{equation}
where $\Mwup$ ($\Mwdown$) is the number of wide roots above (below)
the real line. On the other hand we have  
\begin{equation}\label{eq:Zandzeta}
\begin{split}
	Z_N(+\infty) &= 2\pi(\Imax +\tfrac12) + \zeta_+ \\
	Z_N(-\infty) &= 2\pi(\Imin -\tfrac12) - \zeta_-
\end{split}
\end{equation}
where $\Imax$ ($\Imin$) is the quantum number corresponding to the
largest (smallest) real root or hole and $\zeta_\pm$ is the mod $2\pi$
residue (obviously $|\zeta_\pm|<\pi$)
\begin{equation}\label{eq:zeta}
	\zeta_\pm = [\pm Z_N(\pm\infty) +\pi\delta_S] \text{ mod }2\pi
	= -2\g S +2\pi\left\lfloor\tfrac12 +\tfrac\g\pi\,S \right\rfloor 
\end{equation}
so that 
\begin{equation}\label{eq:Imaxmin}
\begin{split}
	\Imax +\tfrac12 &= +\tfrac12(N+S) +\sign(\pi-2\g)\Mwdown 
	-\left\lfloor\tfrac12 +\tfrac\g\pi\,S \right\rfloor  \\
	\Imin -\tfrac12 &= -\tfrac12(N+S) -\sign(\pi-2\g)\Mwup
	+\left\lfloor\tfrac12 +\tfrac\g\pi\,S \right\rfloor
\end{split}
\end{equation}
Evidently the total number of real zeroes of $ 1+(-1)^S e^{iZ_N(\l)} $ is
\begin{equation*}
\begin{split}
	\MR + \NH  &= \Imax - \Imin +1 + 2\NS  \\ 
 	 &= N +S -2 \left\lfloor \tfrac12+\tfrac\g\pi\,S 
		\right\rfloor  + \sign(\pi-2\g)\,\Mw + 2\NS
\end{split}
\end{equation*}
where $ \NS $ stands for the total number of special roots/holes.

On the other hand we have by definition
\begin{equation*}
	\MR  = N - S - \Mc - \Mw  
\end{equation*}
so that we obtain the following general constraint between the total
number of holes and of complex roots and the U(1) charge $S=N-M$
[$\theta(x)$ is the step function]:
\begin{equation}\label{eq:NHetc}
	\NH -2\NS = 2\left(S - \left\lfloor \tfrac12+\tfrac\g\pi\,S
	\right\rfloor\right) + \Mc + 2\,\Mw\; \theta(\pi-2\g)
\end{equation}
The important fact about eq.\eqref{eq:NHetc} is that it involves only
quantities which are finite as $N\to\infty$. Notice also that $\NH$
turns out to be always even. We may write eq.\eqref{eq:NHetc} in a
different way by introducing $\Mscpri$, the number of self--conjugated
roots of the first class and the ``effective hole number''
\begin{equation}\label{eq:effNH}
	\effNH \equiv \NH -2\NS -2\,\theta(\pi-2\g)\,\Mscpri 
	+ 2 \left\lfloor \tfrac12+\tfrac\g\pi\,S \right\rfloor
\end{equation}
Notice that $\effNH$ indeed stays constant throughout the processes
described in the examples above. We now have the relation
\begin{equation}\label{eq:NHetcI}
	\effNH = 2S + \Mc + 2\,\theta(\pi-2\g)(\Mw-\Mscpri) 
\end{equation}
where evidently $\Mw-\Mscpri$ is the number of wide roots which are
not self--conjugated roots of the first class.   
\bigskip

We may now develop the following interpretation for the complex roots.
First of all we observe that, for any given configuration of quantum
numbers, there exists sufficiently small values of $\g$ such that
$\lfloor\tfrac12+\tfrac\g\pi\,S\rfloor=0$ and
$\NS=\Mscpri=0$. Then $\effNH=\NH$ from eq.\eqref{eq:effNH}
and we see [from  eq.\eqref{eq:NHetcI}] that the holes carry a
U(1) charge $S=\NH/2$ if no complex roots are present, while this
charge is lowered by $1$ for each close root and by $2$ for each wide
root. Next, as $\g$ is raised while keeping all quantum numbers fixed,
$\lfloor\tfrac12+\tfrac\g\pi\,S\rfloor $, $\NS$ and
$\Mscpri$ might become nonzero but $\effNH$ stays fixed.  In
particular, if the $\NH$ holes present for small enough $\g$ are all
well within the real distribution, in the sense that their quantum
numbers are of order $1$ rather than $N$, then these holes stay normal
for any value of $\g$ and are exactly those counted by $\effNH$.  We
anticipate that that these configurations are exactly those relevant
in the continuum limit (see section \ref{continuum})

We may thus draw the conclusion that in the repulsive regime
$\g<\pi/2$ all complex roots (other than self--conjugated roots of the
first class) act as {\em parameters} characterizing the various U(1)
polarization states of $\effNH$ {\em normal holes} well within the
distribution of real roots, which have U(1) charge $1/2$. The allowed
values for the quantum numbers of these complex roots must then be of
order $\NH$ rather than $N$, as can be verified by carefully analyzing
the counting function of BA solutions found numerically.

In the attractive regime $\g>\pi/2$ this interpretation holds for the
close roots only. From eq.\eqref{eq:NHetcI} we see that adding wide roots
does not change the U(1) charge of the $\effNH$ holes since
 $ \theta(\pi-2\g) = 0 $.
Thus, when $\g>\pi/2$, the wide roots appear as {\em independent}
excitations not carrying any U(1) charge and with a phase space of
order $N$.  These picture is made more precise but fully confirmed by
appropriate calculations (including those of the energy--momentum
spectrum) in the $N\to\infty$ limit at fixed $\Th$ (see section
\ref{spec}).

\section{Large $\Th$ limit and relation to the homogeneous six-vertex
model and the XXZ chain}\label{largeTh}

When $\Th$ vanishes the BAE \eqref{eq:baeTh} reduce to those of the
homogeneous six-vertex model and the
XXZ spin 1/2 chain with $2N$ sites. This is the obvious
correspondence. Less obvious and more important for our purposes is
the relation which follows in the limit $\Th\to\infty$ at fixed $N$.
In this limit the BAE split into two separate sets of twisted BAE each
relative to an XXZ chain with $N$ sites. The two sets remain coupled
by the twists, which depend on global properties of the original BA
state as well as on those of the final pair. To see all this we may work
directly with the counting function $Z_N(\l,\Th)$, in which we write
out for clarity the dependence on $\Th$. Notice that this dependence
is both explicit, in the source term proportional to $N$, and
implicit, through the BA roots themselves.

As $\Th\to\infty$ at fixed $N$, almost all the roots separate into two
sets, the left--moving and right--moving seas, having ``center of
mass'' in $+\Th$ and $-\Th$, respectively, and spreading of order
$\log N$ for large $N$. Few roots may stay within finite regions for
special values of $\g$ or sufficiently symmetric root configurations,
as will become clear below.  Let us split the $M$ BA roots into the
two sets $\{\pm\Th+\l_j^\pm;\,j=1,2,\ldots, M^\pm\}$ of the roots
diverging as $\pm\Th$, which we shall call right-- and left--moving,
respectively, plus the $M^0$ roots with a finite limit for $\Th \to \infty $. 
Each 
$\l_j^\pm$ measures the distance of the diverging root from its center
of mass and tends to a finite limit. Then we find for $Z_N(\l,\Th)$:
\begin{equation*}
\begin{split}
	\lim_{\Th\to +\infty} Z_N(\l\pm\Th,\Th) \equiv Z_N^\pm(\l) 
	&= N\phi_{1/2}(\l) -\sum_{k=1}^{M^\pm} \phi_1(\l-\l^\pm_k) \\
	 \pm & (\pi-2\g)(S-S^\pm) \pm 2\pi \left[
		\tfrac14 N + n_{\mathrm{wide}}^\pm \right]
\end{split}
\end{equation*}
where $S^\pm=N/2-M^\pm$ is the U(1) charges of the left and right
seas, respectively, and $n_{\mathrm{wide}}^\pm$ are integer which
depend on the wide roots. Evidently the $M^0$ roots that do not
diverge contribute $S^0=-M^0$ to the total charge, so that
$S=S^-+S^0+S^+$. We anticipate that, while the global charge $S$ is
nonnegative, in certain cases one of the two partial charges $S^\pm$
may take also the value $-1$. For completeness, we also observe that,
due to our choice of logarithmic branches for the function
$\phi_1(\l)$, the integers $n_{\mathrm{wide}}^\pm$ are given by
\begin{equation}\label{eq:npm}
\begin{split}
	n_{\mathrm{wide}}^+ &= \sign(\pi-2\g)[\Mwdown -\Mwdownp]\;,\\
	\quad	
	n_{\mathrm{wide}}^- &= \sign(\pi-2\g)[\Mwup -\Mwupm]
\end{split}
\end{equation}
The quantization rules 
\begin{equation*}
	Z_N^\pm(\l^\pm_j) = 2\pi I^\pm_j
\end{equation*}
now reproduce, upon exponentiation, the twisted BAE of two XXZ
chains.

By direct inspection one finds 
\begin{equation}\label{eq:Zpminf}
	\Om_\pm \equiv Z_N^\pm(\mp\infty) = 
	\pm (\pi-2\g)(S -2S^\pm) \pm 2\pi\ell_{\mathrm{wide}}^\pm
\end{equation}
where
\begin{equation*}
\begin{split}
	\ell_{\mathrm{wide}}^+ &= \sign(\pi-2\g)[\Mwdown -\Mwp]\;,\\
	\quad	
	\ell_{\mathrm{wide}}^- &= \sign(\pi-2\g)[\Mwup -\Mwm]
\end{split}
\end{equation*}
and $\Mwup$ ($\Mwdown$) is the number of wide roots above (below)
the real line. We may also write
\begin{equation*}
	\Om_\pm = \Om
	\pm [ (\pi-2\g)S^0 + 2\pi\ell_{\mathrm{wide}}^0 ]
\end{equation*}
where
\begin{equation}\label{eq:Omega}
	\Om = -(\pi-2\g)\DS + 2\pi\ell_{\mathrm{wide}}
\end{equation}
$\DS=S^+ -S^-$ and 
\begin{equation*}
	\ell_{\mathrm{wide}}^0 = \tfrac12 
	(\ell_{\mathrm{wide}}^+ - \tfrac12 \ell_{\mathrm{wide}}^-)
	\;,\quad  \ell_{\mathrm{wide}} = \tfrac12
	(\ell_{\mathrm{wide}}^+ + \tfrac12 \ell_{\mathrm{wide}}^-)
\end{equation*}
Evidently $\Om $ is the common value of the right asymptote of
$Z_N^-(\l)$ and the left asymptote of $Z_N^+(\l)$ whenever $S^0=0$
(which implies $\ell_{\mathrm{wide}}^0=0$ too). In this case we may
approximate $Z_N(\l,\Th)$ for $\Th$ very large but finite, as
\begin{equation*}
	Z_N(\l,\Th) \simeq Z_N^-(\l-\Th) + Z_N^+(\l+\Th) -\Om
\end{equation*}
Thus $Z_N(\l,\Th)$ has a large almost flat plateau which tends to
$\Om$ as $\Th\to\infty$. This observation applies also when $N$ is
very large, provided $ \Th\gg \log N $, since the size of each sea is of
order $\log N$. In case of non-degenerate configurations $\Om$, the
height of the plateau, fixes the way the quantum numbers $I_j$
relative to real roots and holes are subdivided into those belonging
to the two seas, that is the $I^\pm_j$'s associated to real roots and
holes. In fact, just by definition we must have:
\begin{equation}\label{eq:IOmI}
	2\pi \Immax < \Om  < 2\pi \Ipmin = 2\pi (\Immax +1)
\end{equation} 
On the other hand, if $ S^0 $ is nonzero, then $Z_N(\l,\Th)$ keeps a
nontrivial structure in the neighborhood of the origin; this
structure interpolates between two large plateaus with heights
$\Om_\pm$. For instance, if $S^0=-1$ due to a single real root at
$\l=x_0$, then $Z_N(\l,\Th)$ can be approximated as
\begin{equation*}
	Z_N(\l,\Th) \simeq Z_N^-(\l-\Th) + Z_N^+(\l+\Th)-\Om
	- \phi_1(\l-x_0)
\end{equation*} 
But since $Z_N(x_0,\Th)=2\pi I_0$, where $I_0$ is some integer
plus $\tfrac12\delta_S$, to leading order in $\Th$ we get $\Om=2\pi
I_0$. Comparing with eq. \eqref{eq:Omega} this shows that the real
root may stay at a finite location $x_0$ only for certain special values
of $\g$, unless separately $I_0=0$ and $\Om=0$. 

As a matter of fact the situations characterized by a special value of
$\g$ bear a strict correspondence with those discussed in the example
of section \ref{general}. There we found that for certain special
values of $\g$ one extremal real root was pushed to infinity with
respect to the bulk of the roots, which stayed localized in a finite
region. Here we find the same with respect to a bulk that moves to
$\pm\infty$. The same argument may be repeated also for the threshold
values of $\g$ beyond which special root/hole appear; it suffices to
translate the discussion of section \ref{general} to the inner tails
of the real distributions, that is the right tail of the left sea and
the left tail of the right sea. There is an important difference,
however: in the discussion of section \ref{general} all special values
of $\g$ are functions of the total spin $S$, which is the sum $S^+
+ S^-$ for generic situations; here they are functions of the
difference $\DS=S^+ -S^-$.

So, let us recall that in the example of section \ref{general} we
dealt with the states obtained by removing $S$ real roots from the
ground state distribution. As in section \ref{general} we begin with
the range $0<2S\g<\pi$ for which the configuration is non-degenerate
and then follow its evolution as $\g$ grows.

The simplest case is when $S$ is even and the $2S$ holes are equally
divided into $\NHp=S$ right--moving and $\NHm=S$ left--moving ones (so
that $\Om=0$, see eq.\eqref{eq:Omega}) 
while $\Immax=-1/2$ and $\Ipmin=1/2$ are occupied by
holes. In this case nothing special happens in the central part of the
root distribution when $\Th$ is very large and $\g$ raises beyond
$\pi/(2S)$. This is because the relevant phase space of each sea (the
right tail of the left sea and the left tail of the right sea)
receives from the twist caused by other sea exactly the same
contribution that it looses due to the growth of $\g$. On the other
hand the boundaries of the distribution (the left tail of the left sea
and the right tail for the right sea) may develop special root/holes
as discussed in section \ref{general}. Actually this scenario applies
to the entire range of $\g$ and therefore also applies to the passage through
those points, such as $\g=\pi/(S+1)$, when extremal real roots jumps
to the lines $\Im\l=\pm\pi^2/(2\g)$. This particular class of BA
states is the one treated in ref. \cite{rava}.

Let us now allow $\NHp$ to be different from $\NHm$, but still impose
that the $\Immax$ and $\Ipmin$ are occupied by holes. Then we find
that, as long $2|\DS|\g<\pi$, so that $-\pi(\DS+1)<\Om<-\pi(\DS-1)$,
we have $\NHp=2S^+$ and $\NHm=2S^-$ [recall eq. \eqref{eq:IOmI}]. As
$\g$ exceeds $\pi/(2|\DS|)$, one hole passes from the sea with higher
partial U(1) charge to the other, since the actual $\Immax$ and
$\Ipmin$ increase by one for positive $\Om$ or decreases by one for
negative $\Om$.  When $\g$ is exactly equal to $\pi/(2|\DS|)$ there is
a hole with a finite limit as $\Th\to\infty$, since $2\pi
\Immax=\Om>0$ or $2\pi \Ipmin=\Om<0$. This example shows that even if
$\NH=2S$, $\NHp-\NHm$ may differ from $2(S^+-S^-)$. In particular in
the exchange of one hole from one sea to the other $\DS$ stays
constant as $\NHp-\NHm$ changes by two.  Notice also that a hole may
stay trapped into a finite region only if $\g$ assumes exactly one of
a discrete set of values. Hence we can regard this case as exceptional
and always assume $\NHp+\NHm=\NH$ by continuity.

When either one, or both, of the two quantum numbers closest to
$\Om/(2\pi)$ are occupied by real roots the situation gets more
involved due to the appearance of special root/holes. For instance, if
$\DS$ is a negative even and $\Immax=(|\DS|-1)/2$ is
occupied by a real root, then, as $\g$ exceeds $\pi/(2|\DS|)$, the
plateau $\Om$ gets below $2\pi \Immax$ which therefore
becomes triply degenerate: two new holes, one left--moving and
special, the other right--moving and normal, appear to the right of
the root.  As $\g$ grows further the root and the left--moving hole
``collide'' and then exchange their positions, the hole now being
normal and the root special; when $\g$ reaches $\pi/(|\DS|+1)$ the
root comes back from large negative values of order $-\Th$ to finite
values, so that $S^-$ increases by one while $S^0$ changes from $0$ to
$-1$ and two plateaus are formed, with the left one
$2(\pi-2\g)=2\pi(|\DS|-1)/(|\DS|+1)$ higher than the other; then the
root passes to the right--moving sea when $\g$ exceeds $\pi/(|\DS|+1)$
and we have once more a unique plateau with height $\Om$ larger
than the original one by $2(\pi-2\g)$, in accordance with the fact
that $\DS$ has decreased by two units; next the root ``collides'' with
the right--moving hole and exchanges place with it; at this stage we
have a normal left--moving hole, a special right--moving hole and a
real root associated to the degenerate quantum number
$\Immax$ and the two holes disappears as soon as the new
$\Om$ gets smaller than $2\pi \Immax$, that is when $\g$
exceeds $3\pi/(2|\DS|)$. In figs. 3 and 4 the changing shape of the
central portion of $Z_N(\l)$ is depicted for a special case of the
type just discussed. It is obtained by numerically solving the BAE.  

Another interesting observation concern that case when initially 
$S^+=0$, so that the right sea is a twisted ground state. At the end
we find a state with $S^+=-1$, that is with $M^+=N/2+1$. This is
possible exclusively thanks to the twist induced by the other sea.

We are now in a position to produce  general formulae analogous
to eqs.\eqref{eq:NHetc} and \eqref{eq:NHetcI}. Our purpose is to
suitably relate the plateau heights
$\Om_\pm$ to the hole and complex root content of the
BA state. We proceed as in section \ref{general} and define $\om_\pm$,
with $|\om_\pm|<\pi$, by the relations
\begin{equation*}
	\Om_- = 2\pi(\Immax + \tfrac12) + \om_- \;,\quad 
	\Om_+ = 2\pi(\Ipmin - \tfrac12) - \om_+
\end{equation*}
We may also write 
\begin{equation*}
	\om_\pm = (\mp \Om_\pm +\pi \delta_S ) \text{ mod }2\pi
\end{equation*}
But $\mp \Om_\pm +\delta_S \pi$ is just $2\g(S-2S^\pm)$ plus some
integer multiple of $2\pi$ (see eqs.\eqref{eq:Zpminf} and
\eqref{eq:Omega}), so that we obtain   
\begin{equation}\label{eq:ompm}
	\om_\pm = 2\g(S-2S^\pm) + 4\pi(S^\pm -\hat S^\pm)
\end{equation}
where    
\begin{equation}\label{eq:hatSpm0}
	\hat S^\pm = S^\pm + \tfrac12 \sign(S-2S^\pm)
	\left\lfloor \tfrac12 + \tfrac\g\pi\,|S-2S^\pm| \right\rfloor 
\end{equation}
It is convenient to relate $\hat S^\pm$ to the content of the two seas
in terms of holes, special root/holes and complex roots. To this end
we write  
\begin{equation}
\begin{split}
	\om_- &= \Om_- - 2\pi(\Immax + \tfrac12) \\
	&= \Om_- - 2\pi(\Imin + \MRm + \NHm -2\NSm - \tfrac12)
\end{split}
\end{equation}
and similarly
\begin{equation}
\begin{split}
	\om_+ &= -\Om_+ + 2\pi(\Ipmin - \tfrac12) \\
	&= -\Om_+ + 2\pi(\Imax - \MRp - \NHp +2\NSp + \tfrac12)
\end{split}
\end{equation}
Then using eqs.\eqref{eq:Imaxmin} and the obvious relation
\begin{equation*}
	\MRpm=\tfrac12 N -S^\pm -\Mcpm -\Mwpm
\end{equation*}
yields 
\begin{equation}\label{eq:hatSpm}
	\hat S^\pm =  \tfrac12 [\effNHpm -\Mcpm 
		-2\,\theta(\pi-2\g)(\Mwpm -\Mscpripm)]
\end{equation}
where the effective hole number of each sea reads (compare with
eq.\eqref{eq:effNH}) 
\begin{equation}\label{eq:effNHpm}
	\effNHpm = \NHpm - 2\NSpm  -2\theta(\pi-2\g)\Mscpripm 
	+ \left\lfloor \tfrac12+\tfrac\g\pi\,S \right\rfloor
\end{equation}
We see that $\hat S^+$ and $\hat S^-$ may be regarded as the partial
U(1) charges induced by the holes, with the complex roots of each sea
fixing the global polarization state of the holes (recall the
interpretation in section \ref{general}). $\hat S^\pm$ does not
coincide in general with $S^\pm$ because the splitting of holes in
right-- and left--moving ones is only partially correlated to that of
the roots (recall that $S^\pm$ is {\em defined} by $N/2-M^\pm$), due
to the coupling of the two seas caused by the twists and/or the
possibilities that $\NHp+\NHm<\NH$ or $S^0\neq 0$. It should also be
remarked that, unlike $S^\pm$, $\hat S^\pm$ can be half-odd-integers.
 
The expression \eqref{eq:hatSpm} in terms of holes and complex root
will prove itself very useful in section \ref{conf}. 

\section{The fundamental nonlinear integral equation}\label{NLIE}

In this section we derive a NLIE (nonlinear integral
equation) which is fully equivalent to the original BAE \eqref{eq:baeTh}
for any excited state. The crucial property of such NLIE is that it depends
analytically on the number of sites $N$ 
and does not contain explicitly the real roots 
(which number is of order $N$), but only holes and
complex roots (which number is of order $1$).

In the definition \eqref{eq:Zagain} of the counting function, let us
rewrite the sum over all roots as
\begin{equation}\label{eq:split}
	\sum_{j=1}^M\phi_1(\l-\l_k) = \sum_{j=1}^{\MR+\NH}
	\phi_1(\l-x_j) - \sum_{j=1}^\NH \phi_1(\l-h_j) 
	+ \sum_{j=1}^\MC \phi_1(\l-\xi_j) 
\end{equation}
where we recall that the $x_j$'s are all the the points on the real
lines where $e^{iZ_N(\l)} = (-1)^S$. That is, real roots {\bf and} holes. 
Next we convert the sum over the $x_j$'s into a contour integral
\begin{equation}\label{eq:trick}
	\sum_{k=1}^{\MR+\NH}\phi(\l-\l_k,\g) =\oint_\Gamma
	{{d\mu}\o{2\pi i}}\,\phi_1(\l-\mu)\,
		\der{}\mu \log[1+(-1)^S e^{iZ_N(\mu)}]
\end{equation}
where $\Gamma$ is a closed curve that lies in the analyticity domain
of the integrand and encircles anti-clockwise all the $x_j$'s once. At
this initial stage we prefer to work with derivatives to avoid
worrying about boundary integration constants. We also consider $\l$
in the neighborhood of the real axis. Thus we can write the derivative
of eq. \eqref{eq:Zagain} in the form
\begin{equation*}
	Z_N'(\l) = Nz_0'(\l) - \oint_\Gamma {{d\mu}\o{2\pi i}}\;
	\phi_1(\l-\mu)\, \der{}\mu \log[1+(-1)^S e^{iZ_N(\mu)}]  
\end{equation*}
where the source term reads
\begin{equation}\label{eq:source}
	z_0(\l) = \phi_{1/2}(\l+\Th)+\phi_{1/2}(\l-\Th) 
	+ {1\o N} \sum_{j=1}^\NH \phi_1(\l-h_j) 
		 - {1\o N}\sum_{j=1}^\MC \phi_1(\l-\xi_j)
\end{equation}
We next choose $\Gamma$ to be the boundary of the infinite rectangle
centered in the origin with horizontal sides extending from $-\infty$
to $+\infty$ and vertical sides of length $2\eta$, with
$0<\eta<\tfrac12{\rm min}(\pi,\pi({{\pi}\o {\g}}-1),\s) $, $ \s $ being
the absolute value of the imaginary part of the complex root of
$e^{iZ_N(\l)} = (-1)^S$ closest to the real line. We denote with
$\Gamma_+$ and $\Gamma_-$ the upper and lower halves of $\Gamma$,
respectively. They are both oriented from left to right, so that
\begin{equation*}
	\oint_\Gamma d\mu \ldots = \int_{\Gamma_-} d\mu \ldots  
	- \int_{\Gamma_+} d\mu \ldots
\end{equation*} 
By construction, the counting function enjoys the property
\begin{equation}\label{eq:reflect}
	Z_N({\bar \mu}) - \overline{Z_N(\mu)} =0 \;\text{ mod }2\pi 
\end{equation}
Moreover, by analyticity we have $Z_N(x+iy)=iyZ'_N(x)+...$ so that
$\Im Z_N(\mu)$ has the same sign of $Z'_N(\Re\mu)$ for $\Im\mu$ positive
and small enough. Assuming for the time being that we
are in a non-degenerate case, this suggests  to extract $Z_N(\mu)$ from the
lower half of the integration rectangle by writing 
$1+(-1)^S e^{iZ_N}= e^{iZ_N}[(-1)^S + e^{-iZ_N}]$, so that the
remaining logarithms have no cut ambiguities for $\eta$ small
enough. The necessary corrections for degenerate cases  will be
introduced at the end.

By extracting $Z_N(\mu)$ we obtain a convolution of
$\phi_1'(\l-\mu)$ with $Z_N'(\mu)$ which can be moved to
the real axis thanks to analyticity, yielding
\begin{equation}\label{eq:NLIE0}
\begin{split}
	[(1+K)*Z_N'](\l) &= N z_0'(\l) -i\int_{\Gamma_+} 
	{{d\mu}\o{2\pi}}\,\phi_1'(\l-\mu)\,\der{}\mu 
	\log\left[ 1+(-1)^S e^{iZ_N(\mu)} \right]\\
	& + i\int_{\Gamma_-}{{d\mu}\o{2\pi}}\,\phi_1'(\l-\mu)\,
	\der{}\mu \log\left[ (-1)^S + e^{-iZ_N(\mu)} \right]  
\end{split}
\end{equation}
where $K*f$ stands for the convolution  
\begin{equation*}
	(K*f)(\l) = \intf {{dx}\o{2\pi}}\,\phi_1'(\l-x)f(x)
\end{equation*}
Applying $(1+K)^{-1}$ to both sides of eq. \eqref{eq:NLIE0} and
integrating by parts leads to the following nonlinear
integral equation (NLIE)
\begin{equation}\label{eq:NLIE1}
\begin{split}
	Z_N(\l) &= N z(\l) -i\int_{\Gamma_+} {{d\mu}\o{2\pi}}
	\,G(\l-\mu)\, \log\left[ 1+(-1)^S e^{iZ_N(\mu)} \right] \\
	&+ i \int_{\Gamma_-}{{d\mu}\o{2\pi}}\,G(\l-\mu)\,
	 \log\left[ (-1)^S + e^{-iZ_N(\mu)} \right] 
\end{split}
\end{equation}
where the `dressed' source term $z(\l)$ is by construction the sum of
bulk, hole and complex root contributions (compare with
eq. \eqref{eq:source}), plus an integration constant
$C$ to be determined later
\begin{equation}\label{eq:z}
	z(\l) = ([1+K]^{-1}s)(\l) = \zV(\l) + 
		{1\o N}\left[\zH(\l)+ \zC(\l) \right] +C
\end{equation}
and $G(\l)$ stands for the kernel of the
convolution operator $G=[1+K]^{-1}*K$. Through Fourier
transforms one obtains the following explicit expressions
\begin{align}
	G(\l) &= \intf{{dk}\o{2\pi}}\, e^{ik\l\g/\pi}
	\,{{\sinh(\pi/2-\g)k}\o  
	{2\sinh(\pi-\g)k/2 \,\cosh(\g k/2)}} \label{eq:G} \\   
	\zV(\l) &= \text{gd}(\Th+\l) - \text{gd}(\Th-\l) \\ 
	\zH(\l) &= \sum_{j=1}^\NH \chi(\l-h_j) \label{eq:zH}
\end{align}
where $\text{gd}(x)=\arctan[\sinh(\l)]$ is the so--called hyperbolic
amplitude (or Gudermannian) and $\chi(\l)$ is the odd primitive of
$2\pi G(\l)$. Notice that $\chi(\l)$ coincides with the
soliton--soliton two-body scattering phase shift \cite{mk}. On the
other hand the complex root contribution is different in the two
regimes: when $\g<\pi/2$ we have
\begin{equation}\label{eq:zCR}
	\zC(\l) = - \sum_{j=1}^{\Mc}\chi(\l-c_j) -
	\sum_{j=1}^{\Mw}\phi_{\a/2}(\a(\l-w'_j)) 
\end{equation}
where $ \phi_{\nu}(\l) $ was defined in eq.\eqref{eq:defi}, 
$\a=(1-\g/\pi)^{-1}$ and 
\begin{equation*}
 		w'_j = w_j - i\; \sign(\Im w_j) \tfrac\pi{2}
\end{equation*} 
while when $\g>\pi/2$ we have 
\begin{equation}\label{eq:zCA}
	\zC(\l) = - \sum_{j=1}^{\Mc}\chi(\l-c_j) -
	\sum_{j=1}^{\Mw} \left[ \text{gd}(\l-w''+i\tfrac{\pi^2}{2\g\a}) +
	\text{gd}(\l-w''-i\tfrac{\pi^2}{2\g\a})  \right] 
\end{equation}
where now
\begin{equation*}
		w''_j = w_j - i\; \sign(\Im w_j)\tfrac{\pi^2}{2\g}
\end{equation*}
A rather more compact form of eq. \eqref{eq:NLIE1} is obtained by
choosing $\eta$ to be infinitesimal:
\begin{equation}\label{eq:NLIE2}
	Z_N(\l) = N z(\l) + (G \ast \Q_N)(\l) +C
\end{equation}
where (recall that we assumed $Z'_N(x)>0$, so that the logarithms are
always in their principal determination)
\begin{equation}\label{eq:Zmod2pi}
\begin{split}
	\Q_N(x) &= -i\,\log
	\dfrac{1+ (-)^S\,e^{i Z_N(x+i\e)}}{(-)^S+ e^{-i Z_N(x-i\e)}}\\
	& @>>{\e\to0}>  \{ Z_N(x) + \delta_S \pi\} \text{ mod } 2\pi
\end{split}
\end{equation}
Hence $\Q_N(x)$ jumps downward by $2\pi$ each time $Z_N(x)$ crosses
$2\pi$ times one of the quantum numbers $I_k$, that is when $x$ passes
through a root or a hole, as required. Notice that the assumption that
$Z'_N(x_j)>0$ for any real root or hole $x_j$ could actually be
dropped after the limit $\e\to0$ in \eqref{eq:Zmod2pi}. In practice,
when $\e$ is strictly zero we identify $\Q_N(x)$ as the unique real
function with the following three properties:
\begin{equation*}\label{eq:altdef}
	e^{iQ_N(x)} = (-1)^S e^{iZ_N(x)} \;,\quad |Q_N(x)|\le \pi
	\;,\quad Z_N\to -Z_N \implies Q_N \to -Q_N
\end{equation*} 
Thus $\Q_N(x)$ jumps by $2\pi$ upwards at a special root/hole 
where $Z'_N(x_j)<0$. Of course one may generalize in the same way
the definition of $\Q_N(x)$ in terms of logarithms, provided we
suitably change determination whenever $Z'_N(x_j)<0$.

We may clarify this matter by performing a simple exercise which
backtracks the derivation of the NLIE, eq. \eqref{eq:NLIE2}.
By definition we have, provided $Z'_N(x_j)>0$
\begin{equation*}
	\Q'_N(x) = Z'_N(x) -2\pi \!\!\sum_{j=1}^{\MR+\NH}\delta(\l-x_j) 
\end{equation*}
which inserted into the derivative of eq. \eqref{eq:NLIE2} leads to a
complete cancelation of the hole contribution, so that we obtain
\begin{equation*}
	(1-G)\ast Z'_N = N \zVpri + \zCpri
\end{equation*}
But $1-G=(1+K)^{-1}$ and therefore the last equation is just $(1+K)^{-1}$
applied to the derivative of the original definition of the counting
function, eq. \eqref{eq:Zagain}. 

It should by now be clear how to modify the NLIE in eq. \eqref{eq:NLIE2}
in the degenerate cases. In general we have 
\begin{equation*}
	\Q'_N(x) = Z'_N(x) - 2\pi \!\! \sum_{j=1}^{\MR+\NH}
	\sign(Z'_N(x_j))\, \delta(\l-x_j) 
\end{equation*} 
Let us now denote with $y_k$, $k=1,\ldots,\NS$, the locations of the
special root/holes. Then we also have 
\begin{equation*}
	\Q'_N(x) = Z'_N(x) -2\pi \!\!\sum_{j=1}^{\MR+\NH}\delta(\l-x_j)
	+ 4\pi \!\! \sum_{k=1}^\NS \delta(\l-y_k)
\end{equation*}
Inserting this into eq. \eqref{eq:NLIE2} would not reproduce
$(1+K)^{-1}$ applied to eq. \eqref{eq:Zagain} just because of the last
sum over special root/holes. Hence the convolution of $G$ with
this sum must be subtracted from the source term $Nz'(\l)$ in
eq. \eqref{eq:NLIE2} yielding the modification 
\begin{equation}\label{eq:zS}
	\zH(\l) \longrightarrow \zH(\l) + \zS(\l) \;,\quad
	\zS(\l) = -2\sum_{j=1}^\NS \chi(\l-y_j) 
\end{equation}
on the hole source.  With this simple but crucial change,
eq. \eqref{eq:NLIE2} holds true in general. Of course it could be
recast into the alternative form \eqref{eq:NLIE1} by analytic
continuation and contour deformation.

Let us now take care of the integration constant $C$. One can easily
establish that $C$ must vanish. In fact $Z_N(x)$ is by definition an
asymptotically odd function up to $2\pi$ times an integer, that is
$[Z_N(x)+Z_N(-x)]\to0$ mod $2\pi$ as $x\to\infty$. Thus $(G
\ast\Q_N)(x)$ is asymptotically odd and we need only verify that
$z(+\infty)-z(-\infty)$ has the required extra $2\pi$ times an
integer.

Let us also remark that in the limit $N\to\infty$  at fixed $\Th$ we
have to leading order $Z_N(\l)\simeq N\zV(\l)$. Hence to leading order
the distribution of real roots is exponentially peaked around $+\Th$
and $-\Th$, with a spreading of order $\log N$. This confirm the
anticipation made in section \ref{general} about the size of the
real distribution.

We now observe that the NLIE in eqs.\eqref{eq:NLIE1} and
\eqref{eq:NLIE2} are manifestly analytic in $\l$, so that they may be
used to define $Z_N(\l)$ away from the real axis. This definition may
actually differ, for $|\Im\l|$ large enough, from the original
definition in eq.\eqref{eq:Zagain} due to a different cut structure.
We shall adopt the new definition implied by the NLIE: since in the
NLIE there is only one term of order $N$ and is explicitly known, we
have a better control on the possible values of the quantum numbers
relative to the complex roots. This redefinition implies that these
quantum numbers may be shifted by integers w.r.t. their original
definition.

The NLIE can be analytically continued away from the real axis in a
straightforward manner as long as
$|\Im\l|<\min(\pi,\pi({{\pi}\o{\g}}-1))$. For larger values of $\Im\l$
one must take into account that the first singularity of the kernel
$G(\l)$ can no longer be avoided by deforming the contours
$\Gamma_\pm$. This is because the real line act as natural boundary
for such deformations. For example $\Gamma_+$ cannot be deformed
through the real line because $F(\mu)$ has modulus larger than one
below the real line. Alternatively, one may say that the cuts implied
by the poles of the kernel $G(\l)$ get pinched by the jump
discontinuities of the non-analytic function $Q_N(x)$ when $\Im\l$
reaches $\pm \min(\pi,\pi({{\pi}\o {\g}}-1)) $. Hence the contribution
of such singularity has to be explicitly added via the residue
theorem, resulting in, for $|\Im\l|>\min(\pi,\pi({{\pi}\o {\g}}-1)) $
\begin{equation}\label{eq:det2}
	Z_N(\l) = N z(\l)_\II + \intf\,G(\l-x)_\II \,\Q_N(x)
\end{equation}
where for any function $ f(\l) $ we have defined 
\begin{equation}\label{eq:det2def}
	f(\l)_\II= \begin{cases}
	f(\l)+f(\l-i\pi\,\sign(\Im\l))\; & 0<\g<\pi/2 \\ 
	f(\l) - f(\l-i\tfrac\pi\g (\pi-\g)\,\sign(\Im\l)) \; & \pi/2<\g<\pi 
		   \end{cases}
\end{equation}
Notice that the the second determination $\zV(\l)_\II$ of the
ground--state contribution to the source $z(\l)$ identically vanishes
in the repulsive regime due to the $i\pi$ anti-periodicity of the
$\sinh$ function. Hence the wide roots do not have a phase space of
order $N$ in the repulsive regime, in agreement with the discussion in
section \ref{general}. On the other hand $\zV(\l)_\II$ keeps a
monotonically increasing term of order $N$ in the attractive regime, as
required by the interpretation of the wide roots as independent
excitations of the attractive regime.

We remark that the expression \eqref{eq:det2} implies for
$Z_N(\l)$ a cut structure, in the domain
$|\Im\l|>\min(\pi,\pi({{\pi}\o {\g}}-1)) $, that differs from that of
the original definition \eqref{eq:Zagain}.

It is also quite interesting to observe that the notion of second
determination \eqref{eq:det2def} allows one to write $\zC(\l)$, the
complex root contribution to the source of the NLIE, in a compact form
valid for both regimes:
\begin{equation}\label{eq:zC}
	\zC(\l) = - \sum_{j=1}^{\Mc}\chi(\l-c_j)
		  - \sum_{j=1}^{\Mw}\chi(\l-w_j)_\II 
\end{equation}
In fact the soliton--soliton two-body scattering phase shift
$\chi(\l)$ enjoys the fundamental crossing properties:
\begin{equation*}
	\chi(\l) + \chi(\l-i\pi) = \phi_{\a/2}(\a(\l-i\pi/2))
\end{equation*}
in the repulsive regime $\g<\pi/2$ and 
\begin{equation*}
	\chi(\l) - \chi(\l-i\tfrac\pi\g (\pi-\g)) = 
	\text{gd}(\l-i\pi/2) +\text{gd}(\l+i\pi/2-i\tfrac{\pi^2}{\g})
\end{equation*}
in the attractive regime $\g>\pi/2$ and 

Holes, special root/holes and complex roots which specify the source
term in the NLIE are constrained by the supplementary quantization
rules
\begin{equation}\label{eq:supp0}
\begin{split}
	Z_N(h_j)   &= 2\pi \IH_{\,j} \;,\quad  j=1,2,\ldots,\NH \\ 
	Z_N(y_j)   &= 2\pi \IS_{\,j} \;,\quad  j=1,2,\ldots,\NS \\
	Z_N(\xi_j) &= 2\pi \IC_{\,j} \;,\quad   j=1,2,\ldots,\MC
\end{split}
\end{equation}
Of course the second of these relations is just a repetition of the
first in case of special holes. In case of special real roots the
locations are in general different from those of the holes, but the
corresponding quantum numbers form by construction a subset of those
of the holes (recall in fact that the case of two real roots with the
same quantum numbers is not allowed).  Together with the NLIE the
above quantization rules provide a framework equivalent to the
BAE. The great advantage over the standard algebraic form of the BAE
is the analytic dependence on $N$, which allows to explicitly perform
the continuum limit (recall that by hypothesis the number of holes and
of complex roots stays finite in that limit).  More subtle is the
question concerning the {\em constructive} nature of the NLIE plus
supplementary quantization rules, namely whether they completely
substitute the original definition \eqref{eq:Zagain} and the BAE
\eqref{eq:baeTh}. Our results show that this is indeed so, provided
the proper distinction between normal and special root/holes
is made: however, this distinction is based on the sign of $Z_N'(x)$
in certain points, while $Z_N(x)$ is itself the unknown in the NLIE.
We shall now verify that this is not a loophole, as it might appear at
first sight.

In fact, the NLIE in \eqref{eq:NLIE2} or \eqref{eq:NLIE1} does not
admit a solution for arbitrary choices of the source term $z(\l)$.
Suppose that we take a $z(\l)$ with an almost flat behavior in a large
portion of the real line (in our case these happens for large
$|\Re\l|$). Since the convolution with the exponentially peaked kernel
$G(x)$ acts as a multiple of the identity on constant functions, we
see that in the flat regions the NLIE reduces to a simple algebraic
equation. In our case we obtain from \eqref{eq:NLIE1}, 
for instance as $ x\to\infty $:
\begin{equation}\label{eq:infNLIE}
	X = b + \dfrac{\chi_\infty}\pi ( X \text{ mod } 2\pi )
\end{equation}
Where $X=Z_N(+\infty)+\delta_S \; \pi$, $b=Nz(+\infty)+\delta_S \; \pi$ and 
\begin{equation*}
	\chi_\infty = \pm \chi(\pm\infty) = \pi\intf dx\,G(x)
	={{\pi/2-\g}\o{1-\g/\pi}} 
\end{equation*}
Let us show that eq.\eqref{eq:infNLIE} admits a solution if the constant $b$
falls in some specific intervals fixed by the ratio $\g/\pi$. Since
\begin{equation}\label{eq:n}
	X \text{ mod } 2\pi = X -2\pi n
\end{equation}
for a suitable integer $n$, then eq.\eqref{eq:infNLIE} is solved
immediately by
\begin{equation*}
	X = 2(1-\g/\pi)b - 2n(\pi-2\g)
\end{equation*}
which is consistent with eq.\eqref{eq:n} provided
\begin{equation}\label{eq:rangob}
	|b - 2 \pi n | \le \dfrac\pi{2(1-\g/\pi)} \; .
\end{equation}
For $\g<\pi/2$,
$$
\frac{\pi}2  < {{\pi} \over { 2 ( 1-\g/\pi)}} < \pi \; ,
$$
and we see that the sequence of intervals generated by \eqref{eq:rangob}
varying $n$ does not cover the real line; if $b$ lays in one of the
uncovered segments then eq.\eqref{eq:infNLIE} has no solution. The
special root/holes cure this problem, by preventing
$b=Nz(+\infty)+\delta_S \; \pi$ to enter into the uncovered segments when
$\g$ and/or $\Th$ vary at fixed quantum numbers.
Viceversa, in the attractive regime $\g>\pi/2$, we have
$$
\pi  < {{\pi} \over { 2 ( 1-\g/\pi)}} < \infty
$$
and there
could be several solutions to eq.\eqref{eq:infNLIE} that differ by
integer multiples of $2\pi$. This is connected to the fact that in the
attractive regime wide roots act as independent excitations that do
not affect the phase space for real roots and holes.

Having derived the fundamental NLIE for a generic BA state, we now
turn to the problem of expressing the energy and momentum eigenvalue
directly in terms of the function $Z_N(\l)$. The methods are based
as before on contour integrals.

\section{Energy and Momentum as functionals of $Z_N$}\label{energy}

The energy and momentum of a BA state may be written, taking
eq. \eqref{eq:expEP} into account
\begin{align}
	E\d &= \sum_{j=1}^M \left[ \phi_{1/2}(\Th-\l_j) 
		+\phi_{1/2}(\Th+\l_j)-2\pi \right]	\label{eq:ene}\\
	P\d &= \sum_{j=1}^M \left[ \phi_{1/2}(\Th-\l_j) 
		- \phi_{1/2}(\Th+\l_j) \right]		\label{eq:mom}
\end{align}
The choice of logarithmic branch in the energy ensures that the
contribution of each real root is negative definite. In this way one
finds that the BA states with holes located at the boundaries of the real
distribution carry a large energy of order $\delta^{-1}$ and decouple
in the continuum limit. Hence, as anticipated in sections
\ref{general} and \ref{largeTh}, only states with holes well within
the real distribution will need to be considered in the continuum limit.

As a preliminary step to relate $E$ and $P$ directly
to the counting function we shall first study the quantity
\begin{equation*}
	W(\l) = \sum_{j=1}^M \phi_{1/2}'(\l-\l_j)
\end{equation*}
[As before, we first consider the derivative to avoid worrying with
boundary integration constants].  The sum over roots in this
expression is rewritten as in eq. \eqref{eq:split}. Then the sum over
real roots and holes may be transformed into a contour integral as
done for $Z_N(\l)$, and this integral is then manipulated in much the
same way to obtain the following result
\begin{align}
	W(\l) &= -\sum_{j=1}^\NH \phi_{1/2}'(\l-h_j) +\sum_{j=1}^\MC 
	\phi_{1/2}'(\l-\xi_j) +\intf \phi_{1/2}'(\l-x)\; Z'_N(x)\;dx
	\label{eq:W1}\\ 
	& +i\int_{\Gamma_+} {{d\mu}\o{2\pi}} \,G(\l-\mu)\, 
	\log\left[ 1+e^{iZ_N(\mu)} \right] -i\int_{\Gamma_-}
	{{d\mu}\o{2\pi}} \,G(\l-\mu)\, 
	\log\left[ (-1)^S + e^{-iZ_N(\mu)} \right] \notag
\end{align}
We then use the NLIE to eliminate the term linear in $Z'_N(\l)$ and
add together, for holes and roots, the two fundamental types of
contributions, the direct one already present in eq. \eqref{eq:W1} and
the back--reaction term coming from the NLIE. We obtain, in compact
notation and taking correctly into account the eventual special root/holes
\begin{equation}\label{eq:W2}
	W(\l) = \WV(\l) + \WH(\l) + \WC(\l)  - 
	\intf {{dx}\o{2\pi}}\,\text{gd}'(\l-x)\,\Q_N'(x)
\end{equation}
where, as above, $ \text{gd}(\l)=\arctan[\sinh(\l)]$ and
\begin{align*}
	\WV(\l) &= N\intf dx\,\text{gd}'(x) \; \phi_{1/2}'(\l-x) \\
	\WH(\l) &= -\sum_{j=1}^\NH \text{gd}'(\l-h_j) + 
		2 \sum_{j=1}^\NS \text{gd}'(\l-y_j) 
\end{align*}
while, with the sum still to be performed, 
\begin{equation}\label{eq:WC}
	\WC(\l) = \sum_{j=1}^\MC \phi_{1/2}'(\l-\xi_j) + 
	\intf {{dx}\o{2\pi}}\, \zCpri(x) \; \phi_{1/2}'(\l-x)
\end{equation}
When $\g<\pi/2$ (repulsive regime) one finds that the direct and
back--reaction terms cancel completely out in the case of wide roots,
independently of their location, yielding
\begin{equation}\label{eq:Wcp}
	\WC(\l) = \sum_{j=1}^\Mc \text{gd}'(\l-c_j) 
\end{equation}
Of course the eventual presence of wide roots would anyway keep
affecting $W(\l)$ through $Z_N(\l)$. In the attractive regime
$\g>\pi/2$ there is instead no such cancelation for generic wide pair
positions and the expression for $\WC(\l)$ contains the non-vanishing
extra terms due to the wide pairs. Their explicit form is quite long
and shall not be written out here. We shall come back on this point
later, when we discuss the $L\to\infty$ limit in which these extra
terms simplify considerably.
  
By integrating $W(\pm\Th)$ as written in eq. \eqref{eq:W2} with respect
to $\Th$ one obtains the expressions of the energy and momentum as
functionals of $Z_N$. Some care is required for the integration
constants, since they could in principle be state--dependent. This
step will be performed only after the $N\to\infty$ limit when it is
quite simple.

\section{The continuum limit}\label{continuum}

It is well known \cite{clmtm,clddv,rev} that the $\effNH$ holes
interspersed in the bulk of the distribution of real roots are to be
identified with physical particles which, in the continuum limit
$\d\to 0$ and $\Th\to\infty$ on the infinite lattice, acquire a
relativistic dispersion relation with mass $m
\sim\d^{-1}\exp(-\Th)$. They are the solitons and antisolitons of the
sG model. The locations $h_j$ of the holes plays the role of
rapidities: a hole at $ h $ has energy--momentum $m(\cosh h,\sinh h)$
in the continuum limit.

In the standard light--cone approach\cite{clddv}  one reaches the continuum
limit by keeping only the leading corrections in $1/L$. These are the
terms of order one in the energy and momentum, that is the particle
spectrum with zero energy--momentum density, and the order $1/L$
corrections in the quantization rules for the particle momenta which
yield the S--matrix \cite{kor}. 
In practice we can say that in this continuum
limit one keeps $m^{-1}$, the characteristic size of the excitations,
much smaller than the size $L$ of the system. 

Here we shall instead take the continuum limit keeping all orders in
$1/L$.  This will allow us to study also the case when $mL$ is very
small, which is relevant for the ultraviolet behavior of the sG
model. The objects of interest are the so--called scaling functions,
that is the continuum limit of the quantities $(E-\EV)L$, where $E$ is
the energy of generic excited states and $\EV$ the vacuum energy.

Let us begin with the counting function.  In the continuum limit at
fixed $L$ both $N$ and $\Th$ tend to infinity with the asymptotic
relation
\begin{equation}\label{eq:Th}
	\Th \simeq \log{{4N}\o{mL}}
\end{equation} 
characteristic of a fixed physical mass for the solitons. In this
limit and for any fixed value of $\l$ the vacuum contribution
$N\zV(\l)$ becomes quite simply $mL\sinh\l$. 

As natural in the continuum limit, we consider only BA configurations
which have the vacuum structure for large rapidities $\l$ at small
enough $\g$: there are only real roots and no hole to the far left and
right. Therefore, as $\g$ is raised the first mechanism by which roots
get isolated in the tails is exactly that illustrated in section
\ref{general}: first one special hole is formed simultaneously at both
extremities; then these holes exchange place with the largest and
smallest real roots; finally these tend to infinity and then jump to
the lines $\Im\l=\pm\pi^2/(2\g)$, respectively.  At this stage in each
extremity there is a normal holes if $\g<\pi/2$ or one special hole if
$\g>\pi/2$. Then, as $\g$ raises even further, if $\g<\pi/2$ the
normal holes are first pushed to infinity, so that the vacuum
structure is reproduced and the mechanism can start over again; if
$\g<\pi/2$ instead, the mechanism starts over from the point when each
special hole is about exchange place with the nearest real root.

Thus, for any value of $\g$ we either have the vacuum structure at
each extremity or a symmetric situation such that at each tail we find
either a special root/hole or a self--conjugated root plus a normal or
special hole. All these deformations of the vacuum structure are then
removed to infinity by the continuum limit $N,\Th\to\infty$ and their
contribution to the source $z(\l)$ cancel out by symmetry. Moreover,
since the special root/holes are formed exactly when
$\lfloor\tfrac12+\tfrac\g\pi\,S\rfloor=0$ jumps by one, one finds that
the continuum version of eq.\eqref{eq:effNH} is just
\begin{equation}\label{eq:effNHcont}
	\effNH = \NH -2\NS  
\end{equation}
where  now the  $ \NS $ special root/holes are in the middle of 
the distribution, that is for $ x $ of order $ 1 $.
Likewise, the general relation \eqref{eq:NHetcI} among the numbers of
holes, special root/holes complex roots takes now the form
\begin{equation}\label{eq:NHetcIcont}
	\effNH = 2S + \Mc + 2\,\Mw\; \theta(\pi-2\g)
\end{equation}
We naturally identify $\effNH$ with the number of solitons and
antisolitons.
   
Having established these simple facts, we may write down the NLIE
satisfied by the continuum limit $Z(\l)$ of $Z_N(\l)$, namely
\begin{equation}\label{eq:NLIE4}
	Z(\l) = mL\sinh\l + g(\l) + \intf dx\, G(\l-x)\,\Q(x) 
\end{equation}
where $\Q(x)$ is related to $Z(x)$ as $\Q_N(x)$ to $Z_N(x)$ in
eq. \eqref{eq:Zmod2pi}, that is 
\begin{equation}\label{eq:Q}
	\Q(x) =	-i\,\log
	\dfrac{1+ (-)^S\,e^{i Z(x+i\e)}}{(-)^S+ e^{-i Z(x-i\e)}} 
\end{equation}
and
\begin{equation}\label{eq:g}
	g(\l)= \zH(\l) + \zS(\l) + \zC(\l) 
\end{equation}
is the excitation part of the source. The various contributions of
holes, special root/holes and complex roots are given by
eq.\eqref{eq:zH}, \eqref{eq:zS} and \eqref{eq:zC}.

We recall that the continuum NLIE \eqref{eq:NLIE4} is to be
supplemented by the the quantization rules \eqref{eq:supp0}.
We recall also that the quantization rules for wide pairs require the
second determination of the counting function, which now reads,
according to eqs. \eqref{eq:det2} and \eqref{eq:det2def},
\begin{equation}\label{eq:det2c}
	Z(\l) = mL(\sinh\l)_\II  + g(\l)_\II 
	+ \intf\, G(\l-x)_\II \,\Q(x)
\end{equation}
Let us now perform the continuum limit on the energy--momentum.  We
recall that we need to integrate the quantities $W(\pm\Th)$
w.r.t. $\Th$, where $W(\l)$ is given by eq. \eqref{eq:W2}. Thus we
obtain for the energy and momentum of a generic excited state,
\begin{equation}\label{eq:EPcont}
	E \pm P = \EV + \EHpm + \ESpm  + \ECpm
	\mp  m \intf {{dx}\o{2\pi}}\,e^{\pm x}\,\Q(x) 
\end{equation}
where $\EV$ is the ground state bulk energy
\begin{equation*}
	 \EV = N\d^{-1} \left[-2\pi+\intf d\l\,
           {{\phi_{1/2}({{\pi \l}\o {\g}}+2\Th)}\o{\pi\; \cosh\l }}\right] 
\end{equation*}
$ \EHpm $  and $ \ESpm $ stand for the contributions from holes and special
holes, respectively
\begin{equation}\label{eq:EHES}
	\EHpm  = m\sum_{j=1}^\NH e^{\pm h_j} \;,\quad
	\ESpm = -2 \, m\sum_{j=1}^\NS e^{\pm y_j}
\end{equation} 
$\ECpm$ represents instead the contributions of the complex
roots. In the repulsive regime $\g<\pi/2$ we have quite simply, from
eq. \eqref{eq:Wcp},
\begin{equation}\label{eq:ContrC}
	\ECpm = -m\sum_{j=1}^\Mc e^{\pm  c_j }
\end{equation} 
In the attractive regime $\g>\pi/2$ their expression can be
calculated, with some lengthy algebra, from eq.\eqref{eq:WC} and read 
\begin{equation}\label{eq:ECatt}
	\ECpm  = - m\sum_{j=1}^\Mc e^{\pm  c_j }
	+ m\sum_{j=1}^\Mw \left[ e^{\pm  w_j } +
		e^{\pm (w_j-i\pi \e_j\pi^2/\g)} \right]
\end{equation}
where $\e_j=\sign(\Im w_j)$. It is important to observe that the
contribution of each wide root coincides with the second determination
of the exponential function (see eq.\eqref{eq:det2def} and notice that the
second determination of the exponential vanishes identically in the
repulsive regime) so that we can write for both regimes
\begin{equation}\label{eq:EC}
	\ECpm = -m \, \sum_{j=1}^\Mc e^{\pm c_j} 
	+ m \, \sum_{j=1}^\Mw (e^{\pm w_j})_\II
\end{equation}
Eqs.\eqref{eq:NLIE4} and \eqref{eq:EPcont}, with the supplementary
quantization rules \eqref{eq:supp0} entirely determine the
excited--states scaling functions of the sG model.

The inputs required are the quantum numbers of the holes, the special
root/holes and the complex roots. Notice that the quantum numbers of
the complex roots cannot in general be chosen arbitrarily, since they
are coupled to those of the holes and by the NLIE itself. On the other
hand they stay fixed for all values of $mL$ and may be determined most
conveniently in the infrared limit $mL\to\infty$. This shall be
discussed in the next section; it will become apparent that all
quantum numbers associated to particles (solitons and breathers) may
in practice be freely chosen, while the quantum numbers associated
to the configurations of complex roots which describe the internal
U(1) states of the solitons must be restricted to a limited number of
distinct possible values depending on the particle quantum numbers 
through the higher level BAE.

One should then solve the NLIE \eqref{eq:NLIE4} with the locations of
holes, special root/holes and complex roots as free parameters, to be
fixed later by the supplementary conditions \eqref{eq:supp0}. The
practical feasibility of this is limited to simple enough states, but
the existence of a given procedure for any given state ensures that
each excited--state scaling function can be determined independently
from all other states.

\section{$mL\to\infty$: mass spectrum and S--matrix}\label{spec}

We consider here the limit where the physical size $L$ of the system
diverges: we then expect the interaction among the physical particles
to cease affecting the energy--momentum. Hence the quantities $E-\EV$
and $P$ should approach finite limits equal to a free massive
spectrum. In fact as $mL\to\infty$ it is easy to show that the
nonlinear integral terms in eqs. \eqref{eq:NLIE4} and
\eqref{eq:EPcont} all vanish exponentially fast. Indeed $Q(x)$ is peaked
around $x=0$ as the exponential of an exponential while $ G(\l) $ dies
exponentially for large $ |\Re\t| $.

Thus the leading form of the counting function is 
\begin{equation}\label{eq:Z1}
	Z(\l) = mL\sinh\l + g(\l) 
\end{equation}
In the infinite volume limit $ Z(\l) $ is monotonically increasing since 
the term $  mL\sinh\l $ dominates. Therefore no special root/holes are
present. In the repulsive regime we then obtain for the excitation energy 
\begin{equation}\label{eq:eneasy}
	E-\EV = \sum_{j=1}^\NH m\cosh h_j - 
	\sum_{j=1}^\Mc m\cosh  c_j
\end{equation}
with an analogous expression for the momentum. Notice that the
counting function is certainly monotonic for large $mL$, so that
$\NS=0$ and $\effNH=\NH$. Another important observation now concerns
the close roots $c_j$'s.  Since $\Im\sinh\l>0$ for $ \pi>\Im\t>0 $,
from eq. \eqref{eq:Z1} it is clear that the quantization conditions
for the close roots, $ \exp[iZ( c_j )]=-1 $, require that the $ c_j
$'s move exponentially fast in $mL$ to positions where $ g(c_j) $
develops the right logarithmic singularities. It easy to check that in
these positions pairs of close roots are separated by $ i\pi $ so that
their contribution cancel out in eq. \eqref{eq:eneasy} due to the
anti-periodicity of the cosh function.  Notice that no such driving
exists for wide roots since the second determination $ (\sinh\l)_\II $
vanishes identically when $ \g<\pi/2 $ for the same anti-periodicity
(see eqs.\eqref{eq:NLIE4} and \eqref{eq:det2def}).

The fact that the excitation energy and momentum do not depend at all
on the complex roots confirms their interpretation of quantum numbers
describing the collective internal U(1) states of the solitons. One
must notice indeed that the hole rapidities $ h_j $ are free parameters
at $ L=\infty $, subject only on the restriction of being distinct. It
appear natural, by continuity, to regard the $ \IH_{\,j} $, the quantum
number of the holes, as free distinct half--odd--integers when $ mL $ is
finite.  
  
The situation is more complex in the attractive regime $\g>\pi/2$
when wide roots explicitly enter the expressions for $\ECpm$. The
infrared limit however simplifies the problem, because when $\g>\pi/2$
the second determination $ (\sinh\l)_\II $ does not vanish anymore,
forcing all complex roots, including the wide roots, to fall into
special configurations.

These are of two main types. Given the positive integer $n$ such
that
\begin{equation}\label{eq:interv}
	{n\o{n+1}} < {{\g}\o{\pi}} < {{n+1}\o{n+2}} 
\end{equation}
and defining $\varrho=\pi(\pi-\g)/\g$, there are arrays of up to
$4\lfloor n/2 \rfloor +4$ roots of the form
\begin{equation*}
	(\chi - i l\varrho \;,\; {\bar \chi} + i\pi^2/\g -i l\varrho) \; ;
	\quad l=0,1,\ldots, \lfloor n/2 \rfloor
\end{equation*}
plus complex conjugates (we chose $\Im\chi>0$ here). The array
collapses to one with just $2\lfloor n/2 \rfloor +2$ roots, if $n\leq
2$, whenever $\Im\chi=\pi/2$ and has, quite trivially, one less root
if $\chi$ is self--conjugated, that is
$\Im\chi=\pi^2/(2\g)=(\pi+\varrho)/2$. Most importantly, these arrays
in any case contain two close roots. These are the configurations of
the first type.

The configurations of the second type {\em are made entirely of wide roots
and always have fixed imaginary parts}; they are odd strings of the
form
\begin{equation*}
	\chi \pm i l\varrho \;;\quad \Im \chi=(\pi+\varrho)/2
		\;,\; l=0,1,\ldots,s 
\end{equation*}
and even strings of the form
\begin{equation*}
	\chi  \pm  i l\varrho \;;\quad \Im \chi=\pm \pi/2
		\;,\; l=1,2,\ldots,s
\end{equation*}
where $1 \leq s \leq \pi / \varrho$.

The fundamental difference between the two types of arrays is in their
effect on the counting function and on the total energy--momentum: the
configurations of the first type, which contain close roots, make room
for holes (recall eq.\eqref{eq:NHetcIcont}) but do not affect the
$L=\infty$ energy--momentum, since their contribution to $\ECpm$
vanishes as can be seen from eq. \eqref{eq:ECatt}; on the contrary the
configurations of the second type, which are made solely of wide
roots, do not provide room for holes but do affect in $\ECpm$, which
now contain the extra terms
\begin{equation*}
	\sum_{j=1}^\NB m_{s_j}e^{\pm \Re \chi_j}
\end{equation*}
where 
\begin{equation}\label{eq:espectro}
	m_s=2m\sin(s\varrho/2)
\end{equation} 
is the breather mass spectrum and $\NB$ is the total number of
configurations of the second type. These configurations corresponds to
the breathers, that is the $S=0$ soliton--antisoliton bound states.
On the other hand the configurations of the first type describe the
various polarization states of the soliton--antisoliton system.  We
can now repeat the argument used for the holes in the repulsive
regime: in the attractive regime both the hole parameters (soliton
rapidities) and the locations of second--type configurations (breather
rapidities) are free in the infinite volume. At finite $mL$, their
quantum numbers are free.

The physical S--matrix describing the scattering of the solitons and
their bound states may be calculated directly from eq. \eqref{eq:Z1}:
clearly $-\exp[i g(h_j)]$ is the total phase which a 
physical particle with rapidity $h_j$ accumulates by going around the
circle, since we assumed periodic boundary
conditions. For instance, in a two--hole state with $S=1$ (no complex
roots), we would find, dropping the corrections which vanish in the
$N\to\infty$ limit,
\begin{equation*}
	m\sinh h_1 = \tfrac{2\pi}L \,\IH_{\,1} - \tfrac1L \,\chi(h_1-h_2) 
\end{equation*}
Hence $ \chi(h_1-h_2) $ is the scattering phase--shift between two
solitons or two antisolitons. The rest of the factorizable two--body
S--matrix of the sG model can be reconstructed by considering more
general states with two holes and certain complex roots \cite{kor}.

As is well known \cite{hlba}, hole positions and complex roots for
physical states are connected by the higher level BA equations. This
is a finite set of BA-type equations where the holes act as source
part and the complex roots appear as BA roots. For a given set of
holes, they determine all possible states. The emergence of this
higher level BA structure can be seen quite clearly from
eq.\eqref{eq:Z1}. In the repulsive regime it suffices to evaluate
$Z(\l)$ at the position of each complex root and then sum the result
over the two partners of each close pair, to cancel out the imaginary
parts proportional to $ mL $. There is no need to do this for wide
roots, since the second determination $\sinh(\l)_\II $ of $\sinh(\l)$
vanishes identically. In the attractive regime things are slightly
more complicated: by summing $Z(\l)=2\pi I$ over all the members of an
array of the second kind (which corresponds to a breather) one finds
the quantization rule for the rapidities of the breathers; by summing
$Z(\l)=2\pi I$ over half or over all the members of an array of the
first kind, depending on its size, one finds the higher level BA
relations between the free parameters of the arrays and the rapidities
of the holes.

The exponentially small corrections to the counting function and the
energy--momentum can be calculated by iteration. The two
next--to--leading orders in the case of the ground state were
calculated in this way in ref. \cite{npb}. For the excited states 
one has to take into account that
the supplementary quantization rules eq. \eqref{eq:supp0} have to be
satisfied to the appropriate order in $e^{-mL}$ (notice that the
special configurations of complex roots are valid only to leading
order). We shall not dwell further here on this infrared expansion.

\section{$ mL\to 0 $ : conformal spectrum}\label{conf}

When the dimensionless parameter $r\equiv mL$ is very small the
regions with positive and negative $\l$ where $r\sinh\l \sim 1$ are
very far apart. We then expect that in a generic case the central
portion of $Z(x)$ broadens to a single plateau, or a two--plateau
system, which extends to all $|x|$'s smaller than $\log(2/r)$ and then
rapidly disappears (not necessarily in a monotonic fashion) in favor
of the dominant exponential growth. The two functions  
\begin{equation}
	Z_\pm(\l) = \lim_{r\to 0} Z(\l\pm \log\tfrac{2}{r})
\end{equation}
describing this crossover determine entirely the leading terms as
$r\to 0$ in the energy--momentum.  We shall call $Z_\pm(\l)$ ``kink''
functions, although the ``kink'' terminology applies more precisely to
the function $\Q(x)$ for strictly positive $\e$ (recall
eq. \eqref{eq:Q}): when $Z(x)$ changes exponentially fast $\Q(x)$ dies
like the negative exponential of an exponential (it would oscillate
exponentially fast for zero $\e$). Thus $\Q(x)$ has a central plateau
(or a two--plateau region) of width $\sim -\log (r)$ and height $\om$
with two kink--like drops to zero at the sides of the central region.

Applying the scaling relation to the NLIE satisfied by $Z(\l)$ yields
the two kink equations
\begin{equation}\label{eq:kink}
	Z_\pm(\l) = \pm e^{\pm\l} + g_\pm(\l) + 
	\intf dx\, G(\l-x)\,\Q_\pm(x) 
\end{equation} 
where $ \Q_\pm(x) $ is related to $ Z_\pm(x) $ in the usual way (see
eq. \eqref{eq:Zmod2pi}), while $g_\pm(\l)$ follows from $g(\l)$
through the scaling $\l \to \l \pm \log\tfrac2{r}$. The source
$g_\pm(\l)$ depends on the positions of holes, complex roots and
eventual special root/holes as $r\to 0$.

{F}or instance, the hole parameters $\{h_j\}$ may be divided
in right--moving, left--moving and the rest according to 
\begin{equation*}
	\{h_j\} = \{h_j^{\pm} \pm \log\tfrac2{r}\,,\,h_j^0 \} 
\end{equation*}
where the $h_j^{\pm}$ and $h_j^0$ have finite limits as $r\to
0$. The quantization rules for right--moving and
left--moving holes are written
\begin{equation*}
	Z_\pm(h_j^\pm) = 2\pi \IHpm_{\,j} \;\quad j=1,2,\ldots,\NHpm
\end{equation*}
where by definition $\IHp_{\,j}\ge \Ipmin$ and $\IHm_{\,j}\le \Immax$.
The hole contribution to $g_\pm(\l)$ may now be evaluated to be,
(see eqs.\eqref{eq:zH} and \eqref{eq:g})
\begin{equation}\label{eq:hzH}
	\zH(\l\pm\log\tfrac2{r})  @>>{r\to0}> \zHpm(\l) \pm 
	(\NH-\NHpm) \chi_\infty \;,\quad 
	\zHpm(\l)= \sum_{j=1}^\NHpm \chi(\l-h_j^{\pm})
\end{equation}
where we recall that 
\begin{equation*}
	\chi_\infty = {{\pi/2-\g}\o{1-\g/\pi}} 
\end{equation*} 
Analogous arguments and expressions apply to special root/holes and
the complex roots, as evident from eqs.\eqref{eq:zS} and \eqref{eq:zC}.

As a matter of fact, we can rely on the analysis of the
limit $\Th\to\infty$ performed in section \ref{largeTh}. Indeed,
since $\Th \simeq \log(4N/r)$ in the continuum limit, to reach the
scaling form $Z_\pm(\l)$ of $Z(\l)$ we can follow two different, but
equivalent limiting procedures, namely
\begin{equation*}
	Z_N(\l\pm \Th,\Th) @>>{N\to\infty}> Z(\l\pm \log\tfrac{2}{r}) 
	@>>{r\to 0}> Z_\pm(\l) 
\end{equation*}
and 
\begin{equation*}
	Z_N(\l\pm \Th,\Th) @>>{r\to 0}> Z_N^\pm(\l\pm \log(2N)) 
	@>>{N\to\infty}> Z_\pm(\l) 
\end{equation*}
The intermediate step of the second procedure involves the functions
$Z_N^\pm(\l)$ studied at length in section \ref{largeTh}. The
subsequent scaling as $N\to\infty$ does not affect the
conclusions drawn there, except that it simplifies some formulae due
to the removal of extremal special root/holes and self--conjugated
roots of the first class, as discussed in section \ref{continuum}.

Hence the partial ``hole--induced'' U(1) charges of the
right-- and left--moving sea take the form (see eq.\eqref{eq:effNHpm}
and eq.\eqref{eq:hatSpm})
\begin{equation*}
	\hat S^\pm = \tfrac12 
	[\NHpm - 2\NSpm- \Mcpm - 2\,\theta(\pi-2\g)\,\Mwpm] 
\end{equation*}
We may write also, with obvious notation 
\begin{equation*}
	\hat S^0= \tfrac12 
	[\NHO - 2\NSO- \McO - 2\,\theta(\pi-2\g)\,\MwO] 
\end{equation*}
so that we read from eq.\eqref{eq:NHetcIcont}
\begin{equation*}
	\hat S^+ + \hat S^- + \hat S^0 = S
\end{equation*}
In addition we have $S^+ +S^- +S^0=S$ by definition. One can then verify that
\begin{equation*}
	g_\pm(x) = \zHpm(x) + \zSpm(x) + \zCpm(x)
	\pm 2(S - \hat S^\pm)\chi_\infty \pm 2\pi n_{\mathrm{wide}}^\pm
\end{equation*}
where the integers $n_{\mathrm{wide}}^\pm$ are given in
eq.\eqref{eq:npm}.

{F}rom section \ref{largeTh} we read other important relations like 
\begin{equation}\label{eq:Zpminf1}
	Z_\pm(\mp\infty) = \Om_\pm =
	\pm (\pi-2\g)(S - 2S^\pm) \pm  2\pi\ell_{\mathrm{wide}}^\pm 
\end{equation}
and 
\begin{equation} \label{eq:ompm1}
\begin{split}
	\Q_\pm(\mp\infty) = \mp \om_\pm 
	= 2\g(S-2S^\pm) - 4\pi (\hat S^\pm -S^\pm) 
	= 2\pi(S-2\hat S^\pm) -2(\pi-\g)(S-2S^\pm)
\end{split}
\end{equation}
Evidently $\Om_\pm$ must satisfy the asymptotic form of
eq.\eqref{eq:kink}, that is the ``plateau equation''
\begin{equation}\label{eq:asy}
	\Om_\pm =  g_\pm(\mp\infty) \pm \dfrac{\chi_\infty}\pi \,\om_\pm 
\end{equation}
One easily calculates from eqs.\eqref{eq:zS}--\eqref{eq:zCA}
and \eqref{eq:zS}
\begin{equation*}
	g_\pm(\mp\infty) = \pm 2(S -2\hat S^\pm) \, \chi_\infty
	\pm 2\pi \ell_{\mathrm{wide}}^\pm 
\end{equation*}
One can check that eqs.\eqref{eq:Zpminf1} and \eqref{eq:ompm1} indeed
solve this plateau equation for any value of $\g$ only if
$g_\pm(\mp\infty)$ correctly contains, when required, the contribution
$\pm 2\chi_\infty$ of the special root/hole.

It is convenient to introduce also the function
\begin{equation*}
	\Q_0(x) = -\Om + \sum_{j=1}^{M^0}\phi_1(x-\l_j^0)
\end{equation*}
(recall that $\Om=\tfrac12(\Om_+ + \Om_-)$ and $M^0=-S^0=S^+ +S^- -S$). It
is understood that the $\l_j^0$ are the limit values of the roots with
a finite limit as $r\to 0$, so that $\Q_0(x)$ is is
$r-$independent. In particular we have
$\Q_0(\pm\infty)=-\Q_\mp(\pm\infty)$. We may now write
\begin{equation}\label{eq:dec}
	\Q(x) = \Q_-(x+\log\tfrac2{r}) +\Q_+(x-\log\tfrac2{r}) 
	 + \Q_0(x)  + q(x)
\end{equation}
where $q(\l)$ collects all subleading contributions and vanishes
(albeit non--analytically) as $r\to 0$ uniformly in $x$ (in other
words $q(x) = o(1)$ for any $x$).

Using the decomposition eq. \eqref{eq:dec} we then find for the
energy--momentum
\begin{equation}\label{eq:enemom}
	E \pm P = \EV + \EHpm + \ESpm + \ECpm \mp  \EKpm 
	\mp {m\o{2\pi}}\intf dx\, e^{\pm x} q(x)
\end{equation}
where the hole contribution reads 
\begin{equation*}
	\EHpm = {2\o L} \left[ \sum_{j=1}^\NHpm e^{\pm h_j^\pm}
	+ \dfrac{r}2 \sum_{j=1}^\NHO e^{\pm h_j^0}
	+ \dfrac{r^2}4 \sum_{j=1}^\NHpm e^{\mp h_j^\pm} \right]
\end{equation*}
with a similar expressions for the special root/holes and complex root
contribution $\ESpm$ and $\ECpm$, as can be read from
eqs.\eqref{eq:EHES} and \eqref{eq:EC}. The kink
contribution reads
\begin{equation*}
\begin{split}
	\EKpm &=  {m\o{2\pi}}\intf dx, e^{\pm x} \left[
	\Q_-(x+\log\tfrac2{r})+\Q_+( x-\log\tfrac2{r})+\Q_0(x) \right] \\
	&= {1\o{\pi L}} \intf dx\, e^{\pm x} \Q_\pm(x)
	+ {m\o{2\pi}} \intf d\t\, e^{\pm x} 
	\left[ \Q_0(x) - \Q_0(\pm\infty) \right]  \\
	&~~~ + {{m^2L}\o{4\pi}} \intf d\t\, e^{\pm x} 
	\left[ \Q_\mp(x) - \Q_\mp(\pm\infty) \right] \\
	&= {1\o{\pi L}}\intf dx\, e^{\pm x}  \left[\Q_\pm(x) 
	\mp \dfrac{r}2 \Q'_0(x) \mp \dfrac{r^2}4 \Q'_\mp(x) \right]  
\end{split}
\end{equation*}
The last integral iin eq.\eqref{eq:enemom} with $q(\l)$ contains
corrections vanishing non--analytically as $r\to 0$. Their explicit
calculation can be done with Wiener--Hopf techniques as in
ref.\cite{npb}. On the other hand the terms of order $L^{-1}$, which
are those relevant from the conformal theory viewpoint, may be found
without even solving the kink equations. The basic ingredient is the
following lemma:

\medskip

{\bf LEMMA}. Assume that $f(x)$ satisfies the nonlinear integral
equation
\begin{equation}\label{eq:nlnr}
	-i\log f(x)=\varphi(x)+ \intf dy\,G(x-y)\,F(y)          
\end{equation}
where $F(x)=2\, \Im\log[1+f(x+i\e)]$, $\varphi(x)$ is real and 
$G(x)=G(-x)$ is real too, with bounded integral ($L^1$) and peaked
around the origin. Eq.(\ref{eq:nlnr}) tells us that $  f(x) $ has unit
modulus for real $ x $. In addition, we assume that when $f(x+i \e)$
is real then $f(x+i \e) > - 1$. 

Then,
\begin{equation}\label{eq:lem}
	\intf dx\,\varphi'(x)\,F(x) =  -2\, \Re\; \int_{\Gamma} {{du}\o u}
	\log(1+u) - \frac12 \left[F_+^2 - F_-^2 \right]\intf dx\,G(x) \; 
\end{equation}
where $F_\pm=F(\pm\infty)$ and $\Gamma$ is any contour in the complex
$u-$plane that goes from $f_-=f(-\infty)$ to $f_+=f(+\infty)$
(avoiding by hypothesis the logarithmic cut from $-\infty$ to $-1$).

This lemma can be proved as follows.
Replacing $ \varphi'(x) $ in the l. h. s. through
eq. (\ref{eq:nlnr}) yields
\begin{equation*}
\begin{split}
	\intf dx\,\varphi'(x)\,F(x) &= 2\; \Im \intf dx\, \left[ 
	-i{{d}\o{dx}} \log f(x) - \intf dy\,G'(x-y)\,F(y) \right]
	\times \\  &\times \log[1+f(x+i\e)] \\
	&= -2\, \Re\; \int_\Gamma {{du}\o u} \log(1+u) - 
	\intf dx\,\intf dy\, F(x)\, G'(x-y)\,F(y) 
\end{split}
\end{equation*}
One must be careful now with the double
integral, which would seem to vanish by symmetry under
$x\rightleftharpoons y$. In fact one is allowed to interchange the two
integrations only if uniform convergence holds. This is however not
true if $\varphi(x)$ and therefore $F(x)$ do not vanish at
infinity. Proceeding with due care one finds, for $a<b$,
\begin{equation*}
\begin{split}
	I(a,b) &= \int_{-a}^a dx\,\int_{-b}^b dy\,F(x)\,G'(x-y)\,F(y)\\
	&= \int_{-a}^a dx\,F(x)\left[ \int_a^b dy\,G'(x-y)\,F(y)
	+  \int_{-b}^{-a} dy\, G'(x-y)\,F(y) \right]
\end{split}
\end{equation*}
Hence, upon integration by parts and letting $b>a\to\infty$
\begin{equation*}
\begin{split}
	I(a,b) &\simeq F_- \int_{-a}^a dx\, [G(x+b)-G(x+a)] 
	+ F_+ \int_{-a}^a dx\, [G(x-a)-G(x-b)] \\    
	&\simeq \tfrac12 \left[F_+^2 - F_-^2 \right] \intf dx\,G(x)
\end{split}
\end{equation*}
Therefore we obtain the identity \eqref{eq:lem}.
\bigskip

We are now in the position to explicitly calculate the conformal
dimensions encrypted in the order $L^{-1}$ term of the
energy--momentum. To make the notation lighter, we shall restrict our
attention to $E+P$. The other chirality $E-P$ follows by applying the 
appropriate symmetries.
We need to evaluate the quantity
\begin{eqnarray}\label{eq:A}
	A_+ &\equiv& \lim_{r\to 0} \tfrac12\, L (\EHp + \ESp
	+ \ECp - \EKp) \cr \cr
	&=&  \sum_{j=1}^\NHp e^{h_j^+} -2 \, \sum_{j=1}^\NSp e^{y_j^+} 
	\!\! -\sum_{j=1}^\Mcp e^{c_j^+} + \sum_{j=1}^\Mwp (e^{w_j^+})_\II
		- {1\o{2\pi}}\intf dx\, e^x \Q_+(\t) 
\end{eqnarray}

We know that $Z_+(\t)$ solves the equation
\begin{equation*}
	Z_+(\l) = \varphi_+(\l) + \intf dx\, G(\l-x)\,\Q_+(x)
\end{equation*} 
with $ \varphi_+(\l)=e^\l + g_+(\l) $ and that the various unknown
parameters $h_j^+$, $y^+$, $c_j^+$, $w_j^+$ are quantized according to
\begin{equation*}
	Z_+(h_j^+) = 2\pi \IHp_{\,j} \;,\quad 
	Z_+(y_j^+) = 2\pi \ISp_{\,j} \;,\quad 
	Z_+(c_j^+) = 2\pi \Icp_{\,j} \;,\quad
	Z_+(w_j^+) = 2\pi \Iwp_{\,j} 
\end{equation*}
Summing $Z_+(h_j^+)$ over $j$ now yields the relation
\begin{equation}\label{eq:IH}
	2\pi \IHp \equiv 2\pi \sum_{j=1}^\NHp \IHp_{\,j} 
	= \sum_{j=1}^\NHp \left[ e^{h_j^+} + g_+(h_j^+)\right] 
	+ {1\o{2\pi}}\intf dx\,\zHppri(x)\,\Q(x)
\end{equation}
where we used the relation between $\zH$ and $\chi$ (see
eq. \eqref{eq:zH}). Next we sum $Z_+(h_j^+)$ and $Z_+(c_j^+)$
over special root/holes and close roots, respectively; we obtain  
in a closely parallel way
\begin{equation}\label{eq:IS}
	4\pi \IS \equiv 4\pi \sum_{j=1}^\NSp \ISp_{\,j} 
	= 2\sum_{j=1}^\NSp \left[ e^{y_j^+} + g_+(y_j^+) \right]
	- {1\o{2\pi}}\intf dx\, \zSppri(x) \,\Q_+(x)
\end{equation}
\begin{equation}\label{eq:Ic}
	2\pi \Icp \equiv 2\pi \!\! \sum_{j=1}^\Mcp \Icp_{\,j} 
	= \sum_{j=1}^\Mcp \left[ e^{c_j^+} + g_+(c_j^+) \right]
	- {1\o{2\pi}}\intf dx\, \zcppri(x) \,\Q_+(x)
\end{equation}
In the case of wide roots we must recall that the second
determination has to be used, according to eqs. \eqref{eq:det2c} and
\eqref{eq:det2def}). Thus we have 
\begin{equation*}
	Z_+(w_j^+) = e^{w_j^+}_\II +  g_+(w_j^+)_\II + 
	\intf dx\, G(w_j^+ -x)_\II\,\Q_+(x) = 2\pi \Iwp_{\,j}
\end{equation*}
and summing $Z_+(w_j^+)$ over $j$ now gives 
\begin{equation}\label{eq:Iw}
	2\pi \Iwp \equiv 2\pi \!\! \sum_{j=1}^\Mwp \Iwp_{\,j} 
	= \sum_{j=1}^\Mcp  [e^{w_j^+}_\II +  g_+(w_j^+)_\II]
	- {1\o{2\pi}}\intf dx\, \zwppri(x) \,\Q_+(x)
\end{equation}
where we have used the relation
\begin{equation*}
	\zwppri(x) = -2\pi \!\!\sum_{j=1}^\Mwp  G(x-w_j)_\II
\end{equation*}
which follows from eqs.\eqref{eq:zC}.

We may now use eqs. \eqref{eq:IH}--\eqref{eq:Iw} to eliminate the sum
over exponentials in eqs.\eqref{eq:A}; at the same
time the derivative of the complete source term is reconstructed in
the integral with $\Q_+(x)$. Thus we obtain
\begin{equation}\label{eq:A1}
\begin{split}
	A_+ &= 2  \pi\,(\IHp -2\ISp -\ICp) - 
	{1\o{2\pi}} \intf dx \,\varphi_+'(x)\, \Q_+(x)  + \Sigma_+  \\  
	\Sigma_+ &= -\sum_{j=1}^\NHp g_+(h_j^+) +2\sum_{j=1}^\NSp g_+(y_j^+)
	 + \!\!\sum_{j=1}^\Mcp g_+(c_j^+)
	 + \!\!\sum_{j=1}^\Mwp g_+(w_j^+)_\II 
\end{split}
\end{equation}
Now, by exploiting the oddness of $\chi(\t)$ and $\phi_\nu(\l)$,
one may verify that all terms in $\Sigma_+$ which depend explicitly on
the positions of holes, special root/holes and complex roots cancel
out completely, leaving behind only constants, namely
\begin{equation}\label{eq:smame}
	\Sigma_+ = -4 \hat S^+ (S - {\hat S}^+) \,\chi_\infty 
	+ 2\pi q_{\mathrm{wide}}^+ 
\end{equation}
where $q_{\mathrm{wide}}^+$ is a rather involved integer or
half--odd--integer which vanishes when no wide roots are present. Its
explicit form is more conveniently determined case by case. The integral in
eq. \eqref{eq:A1} is computed directly from the lemma, upon the
identifications
\begin{equation*}
	f_\pm = \exp[Z_+(\pm\infty+i\e)] \;,\quad 
	F_\pm = \Q_+(\pm\infty) = 2\,\Im\log(1+f_\pm) 
\end{equation*}
The $\e$ is important at $+\infty$, where $Z_+(\t)$ diverges
exponentially; hence one must let $\e\to0$ {\em after} the
$\t\to+\infty$ limit. On the other hand $Z_+(\t)$ tends to a constant
as $\t\to-\infty$, so that we can set $\e=0$ for $f_-$.  We have
therefore $f_+=F_+=0$, $f_-=e^{i\om}$ and $F_-=\om$ and so, 
Recalling that the integral of $ G(\t)$ over the real axis is just
$\chi_\infty/\pi$,
\begin{equation*}
	\intf dx \,\varphi_+'(x)\, \Q_+(x) = -2\Re \int_\Gamma 
	{{du}\o u} \log(1+u) +\dfrac{\om_+^2 \chi_\infty}{2\pi}
\end{equation*}  
We now choose $\Gamma$ to be the union of the arc of circle from 
$e^{i\om_+}$ to $1$ and the straight segment from $1$ to $0$, so that
\begin{equation*}
	-2\Re \int_\Gamma {{du}\o u} \log(1+u) =
	2\int_0^1 {{du}\o u} \log(1+u) -2 \int_0^{\,\om_+} d\a\; 
	\Im \log(1+e^{i\a}) = \dfrac{\pi^2}6 - \dfrac{\om_+^2}2 
\end{equation*}
and
\begin{equation}\label{eq:integrfi}
	{ 1 \o {2 \pi}}\intf dx \,\varphi_+'(x)\, \Q_+(x) = 
	\dfrac{\pi}{12} - \dfrac{\om_+^2}{8\pi\,(1-\g/\pi)}
\end{equation}
Using eq.\eqref{eq:ompm1} and extending the derivation to the negative
chirality in the obvious way, finally yields
\begin{equation}\label{eq:A2}
\begin{split}
	\frac1{2\pi}\;
A_\pm &=  -\tfrac1{24} + \dfrac{S^2}{4(1-\tfrac\g\pi)}
	+\dfrac14 (1-\tfrac\g\pi)(S-S^\pm)^2 -\tfrac12 S^2 
	+(S-2\hat S^\pm)(S^\pm-\hat S^\pm) \\ 
	& \pm (\IHpm -2\ISpm -\ICpm +q_{\mathrm{wide}}^\pm)
\end{split}
\end{equation}
This result can be rewritten more conveniently as
\begin{equation}\label{eq:Afin}
	\tfrac12 (E -\tfrac12 \EV \pm P) \simeq 2\pi L^{-1} 
	\left[-\tfrac1{24} +\Delta_{\text{sG}}^\pm + n_\pm \right]
\end{equation}
where
\begin{equation}\label{eq:confD}
	\Delta_{\text{sG}}^\pm  =  
	\dfrac{[S - (1-\g/\pi)(S-2S^\pm)]^2}{4(1-\g/\pi)} 
\end{equation}
and
\begin{equation}\label{eq:secoper}
	n_\pm = \pm (\IHpm -2\ISpm -\ICpm +q_{\mathrm{wide}}^\pm) 
		- \hat S^\pm (S +2S^\pm -2\hat S^\pm)
\end{equation}
Eqs. \eqref{eq:Afin}-\eqref{eq:secoper} display the spectrum of a
conformal field theory with central charge $c=1$. The excitations
spectrum corresponds to a Coulomb gas and represent the conformal
dimensions at the ultraviolet fixed point of the operators which
interpolate each given state [see section IX]. $\Delta_{\text{sG}}^\pm$ are to
be identified with the conformal dimensions of primary
operators. These operators are labeled by the U(1) charge $S$ and the
partial U(1) charge $S^\pm$. In eq.\eqref{eq:Afin} we see the integers
$n_\pm$ added to $\Delta_{\text{sG}}^\pm$ inside the bracket. It can be
verified that they are always nonnegative; when positive, they
indicate that the corresponding Bethe state is associated to a
secondary conformal operator.  

We consider now some relevant examples.

\begin{flushleft}
$\bullet$ {\bf States with no complex roots}
\end{flushleft}

We begin with the states without complex roots
of any type, as done already in sections \ref{general} and
\ref{largeTh}.  For $S$ even and $\g$ small enough, we have $\NS=0$,
$\NH=2S$, $S^++S^-=S=\hat S^++\hat S^-)$ and 
$\hat S^\pm=S^\pm=\NHpm/2$.  Recall in fact eq.\eqref{eq:hatSpm0}:
\begin{equation}\label{eq:hatSpm0'}
	\hat S^\pm = S^\pm + \tfrac12 \sign(S-2S^\pm)
	\left\lfloor \tfrac12 + \tfrac\g\pi\,|S-2S^\pm| \right\rfloor 
\end{equation}
Now the smallest value of $\pm\IHpm$, for a given value of
$\NHpm=2S^\pm$ is attained when the $\IHpm_{\,j}$ are all consecutive
half--odd--integers starting from $\Ipmin=(1-\DS)/2$ and
$-\Immax=(1+\DS)/2$, respectively. In this way a single sequence of
holes without interruptions is formed and we find
\begin{equation*}
	\pm\IHpm = \sum_{j=1}^\NHpm \left[
		\tfrac12 (1 \mp\DS) +j-1 \right] = S^\pm S
\end{equation*}
so that we find $n_\pm=0$, which shows that these states are
interpolated by primary operators. We see that the secondary operators
in these conformal towers correspond to the states when the holes are
arbitrarily distributed and their sequence has gaps. Thus the full
conformal tower reproduces the phase space of the $2S$ holes.

At larger values of $\g$ we might have $\hat S^\pm\neq S^\pm$. In the
case of the primary states this happen because $\hat S^\pm$ changes
while $S^\pm$ stays fixed. Hence $\Delta_{\text{sG}}^\pm$ does not
change while $S$ gets divided in different ways into $\hat S^\pm$ for
different values of $\g$. At the special values of $\g$ where $\hat
S^\pm$ jumps we have $\hat S^0=1/2$: one hole is passing from the sea
with higher U(1) charge to the other (recall the discussion in section
\ref{largeTh}). In any case one finds that $n_\pm$ stays constant an
equal to $0$ for all $\g$: in fact, if $\hat S^+=S^+-1/2$ and $\hat
S^0=1/2$, for instance, then $\IHp\to\IHp-\Ipmin=S^+S-(1-\DS)/2=\hat
S^+(S+2S^+-2\hat S^+)$, while the negative chirality sea is not
modified at all.

It is easy to check that the same conclusions apply when a special
root/hole is formed by raising $\g$ when $\Immax$ and/or $\Ipmin$ is
occupied by a real root (notice that this implies a secondary state):
the conformal dimensions of the state do not change. On the other
hand, when the real root passes from one sea to the other as $\g$
crosses one of a critical set of rational values (see again 
the section \ref{largeTh}), $S^\pm$ do change and the state switch
from a conformal tower to another. Notice that exactly at the critical
value $S^0=-1$ and there are two plateaus in $Z(\l)$.

When $S$ is odd (still no complex root) there is the new possibility
that $S^0=-1$ for all $\g$ with $S^\pm=(S+1)/2$.  At the quantization
value $Z(x)=0$ are associated a special real root and two normal
holes if $\g<\pi$ and just a normal real root if $\g>\pi$.  Indeed we
have $\NHpm=2\hat S^\pm = 2S^\pm - \lfloor\tfrac12+\tfrac\g\pi\rfloor$
by eq.\eqref{eq:hatSpm0'}. However, even if the sequence of holes is
interrupted by the real root at $Z(x)=0$, these $\g-$generic
two--plateau configurations contain a primary state. In fact, when the
holes are maximally packed around the origin, since
$\Ipmin=-\Immax=\lfloor\tfrac12+\tfrac\g\pi\rfloor$ we find
\begin{equation*}
	\IHpm = \sum_{j=\Ipmin}^S j =  \sum_{j=1}^S j = 
	\tfrac12 S(S+1) = \hat S^\pm (S +2S^\pm-\hat S^\pm)
\end{equation*}
and eq.\eqref{eq:secoper} gives $n_\pm=0$.

\begin{flushleft}
$\bullet$ {\bf Zero charge states }
\end{flushleft}
 
Let us consider now further illuminating examples. Let us begin with
the BA states with two holes and $ S=0 $ that in the large volume limit
describe the scattering of a soliton--antisoliton pair. The
antisymmetric state contains two holes and one close pair in both
regimes. The symmetric state contains instead two holes and one
self--conjugated root in the repulsive regime and two holes and one
degenerate array of the first kind based on a self--conjugated root
(see section \ref{spec}) in the attractive regime. In any case one
finds from eq.\eqref{eq:NHetcIcont} that $S=0$, as required. The
quantum number of each complex roots is entirely fixed by the number
of holes through the higher level BA and takes therefore a unique
value.

We now let $r\to 0$ keeping all quantum numbers fixed and assuming for
simplicity that a unique plateau is formed. The two holes and the
complex roots are then either all right--moving or all left--moving,
so that $\hat S^\pm=S^\pm=0$.  One finds that the sum $\ICpm$ over all
complex roots identically vanishes (taking into account also the
integer or half--odd--integer $q_{\mathrm{wide}}^\pm$ in
eq.\eqref{eq:secoper} when there are wide roots). Hence we have
$\Delta_{\text{sG}}^\pm=0$ with either $n_+=0$ and $n_-=\IHm\ge 2$ or
$n_-=0$ and $n_+=\IHp\ge 2$.

Next let us consider the states with only wide roots and no holes in
the attractive regime. This are the breather states (notice that the
string--like configurations proper of the large volume limit
$r\to\infty$ of section \ref{spec} are largely deformed in the
opposite limit). Also these states have all $\Delta_{\text{sG}}^\pm$
since all the various U(1) charges identically vanish. One also finds 
$n_\pm\ge 0$.   

Therefore it would appears that the states with zero charge all
correspond to descendants of the unit operator from the conformal
viewpoint. On the other hand we should not forget that
eq.\eqref{eq:Afin} represent only the leading term in the $r\to 0$
limit. In the subleading corrections there should be differences among
the various zero charge states that highlight the different ultraviolet
properties of the operators that interpolate such states.  In
particular the states with only one self--conjugated root in the
attractive regime are those of the lightest breather, that is the
fundamental boson interpolated by the sG field itself. Since the
ultraviolet fixed point of the sG model is the free massless boson
field theory, the two--point function of the sG field has a
logarithmic singularity at short distances. This has to appear as 
$\log(2/r)$ correction in the scaling functions of the lightest
breather. We remand a detailed analysis of these aspects of the
breather states to further studies.

Finite size corrections in lattice models
have been computed in ref.\cite{klupe} using
related but somehow different methods. Our derivation of the NLIE and
the calculational methods based on it are simpler and apply to a wider
set of models. Moreover, we better 
control the constant pieces that yield the descendant fields states.

\section{The Coulomb gas and Duality Symmetry}

The conformal dimensions for a Coulomb gas (central charge $c=1$)
take the form \cite{kada}
\begin{equation}\label{eq:confCG}
	\Delta_{e,\,m}(R) = {1 \o {2 R^2}} \left( {e\o2} + m R^2 \right)^2
\end{equation}
where $e,\,m\in\mathbb Z$ stand for the ``electric'' and
``magnetic'' charges and $ R $ for the compactification radius.

Notice that the spectrum  \eqref{eq:confCG} is invariant under
`electromagnetic' duality \cite{kada}
\begin{equation}\label{eq:dual}
	e \leftrightarrow 2 m \quad , \quad R  \leftrightarrow 1/R 
\end{equation}
This duality in fact maps conformal dimensions with {\bf even} $ e
$. It is then an invariance for the subset of primary fields $ (e,m) $
with even $ e $.

More generally, for any natural number $ K $ we have the following 
$K$-duality invariance,
\begin{equation}\label{eq:dualK}
	\Delta_{e,m}(R) =\Delta_{2mK,\, e/(2K)}(K/R)
\end{equation}
This is an endomorphism for conformal states with an `electric' charge $ e $
which is a multiple of $ 2 K $. For $ K = 1 $ we recover the duality
defined by eq.\eqref{eq:dual}.

A look to our results for the sine-Gordon model
[eq.\eqref{eq:confD}] shows that the compactification radius has
the value
\begin{equation}\label{eq:compaR}
	R= \sqrt{2  (1-\g/\pi)} = {{\b} \o {\sqrt{4\pi}}}
\end{equation}
That is, 
\begin{equation}\label{eq:cDsG}
	\Delta_{\text{sG}}(R) ={1 \o {2R^2}}\left[S + \tfrac12 R^2 \DS\right]^2 
\end{equation}
Hence the ``electric'' and ``magnetic'' charges for sine-Gordon are
identified as
\begin{equation*}
	e = 2 S =2(S^+ + S^-) \;, \quad m = \tfrac12\DS = \tfrac12(S^+ - S^-)
\end{equation*}
in terms of the two partial U(1) charges.

Notice that $ 0 \leq R \leq \sqrt2 $ since $ 0 \leq \g \leq \pi $. 
We see that $ K/R $ does not always belong to this interval. For $ K = 1 $, 
\begin{equation*}
	{1 \o {  \sqrt2}} \leq  { 1 \o R } \leq \infty
\end{equation*}
and there is the nontrivial overlap $ ({1 \o {  \sqrt2}},  \sqrt2) $
between the allowed values for $ R $ and $ 1/R $. In particular,
the invariant point of the duality mapping, $ R = 1 $, is within such interval.
$ R = 1 $ corresponds to the free field point $ \g = \pi/ 2 $ and $
\b^2 = 4 \pi $. 

$ R =  \sqrt2 $ corresponds to the rational limit of
the six-vertex model ($ \g = 0 $) and to the strong repulsive limit
of sine-Gordon ($ \b^2 = 8 \pi $), where it becomes strictly
renormalizable and equivalent to the SU(2) Thirring model. 
 
$ R ={1 \o {  \sqrt2}} $  corresponds to $  \g = 3\pi/ 4 $ in the
attractive regime. This is the threshold for the third
soliton-antisoliton bound state  ($ s = 3 $ in  eq.\eqref{eq:espectro}).

For $ K = 2 $, only the fixed point $ R =  \sqrt2 $ ($ \g = 0 $) is
mapped inside the allowed interval. This duality corresponds to the
weak-strong coupling mapping in the sinh-Gordon and
Toda field theories discussed in ref.\cite{toda}. [Notice that $ \beta
$ becomes $ i  \beta $ for sinh-Gordon and Toda theories yielding a
different duality structure.]

For $ K > 2 , \; K/R $ is always outside the interval $ (0,  \sqrt2) $.

It is interesting to compare the conformal dimensions \eqref{eq:confD}
for the ultra-relativistic (ultraviolet) limit of the sine-Gordon
theory with those for the low energy limit (infrared) of the
six-vertex model \cite{rev,seisV}. These can be written
\begin{equation}\label{eq:conf6V}
\begin{split}
	\Delta_{\text{6V}} &=
	{1\o{4(1-\g/\pi)}}  \left[ \DS + (1-\g/\pi) S \right]^2 \\
	&= {1\o{2R^2}}  \left[ \DS + \tfrac12 R^2\, S \right]^2
\end{split}
\end{equation}
where $ S^\pm \equiv \frac12 ( S \pm \DS ) $ now is the contribution to the
third component of the spin  
due to the right and left tail of the BA distribution, respectively.
We see that the $ \Delta_{\text{6V}} $ and the  $ \Delta_{\text{sG}}$ are 
connected by the exchange $ S \leftrightarrow \DS$ between U(1) charge
and chiral U(1) charge. Such exchange is equivalent to make $ R
\leftrightarrow R^{-1} $.

\begin{center}
       \epsfig{file=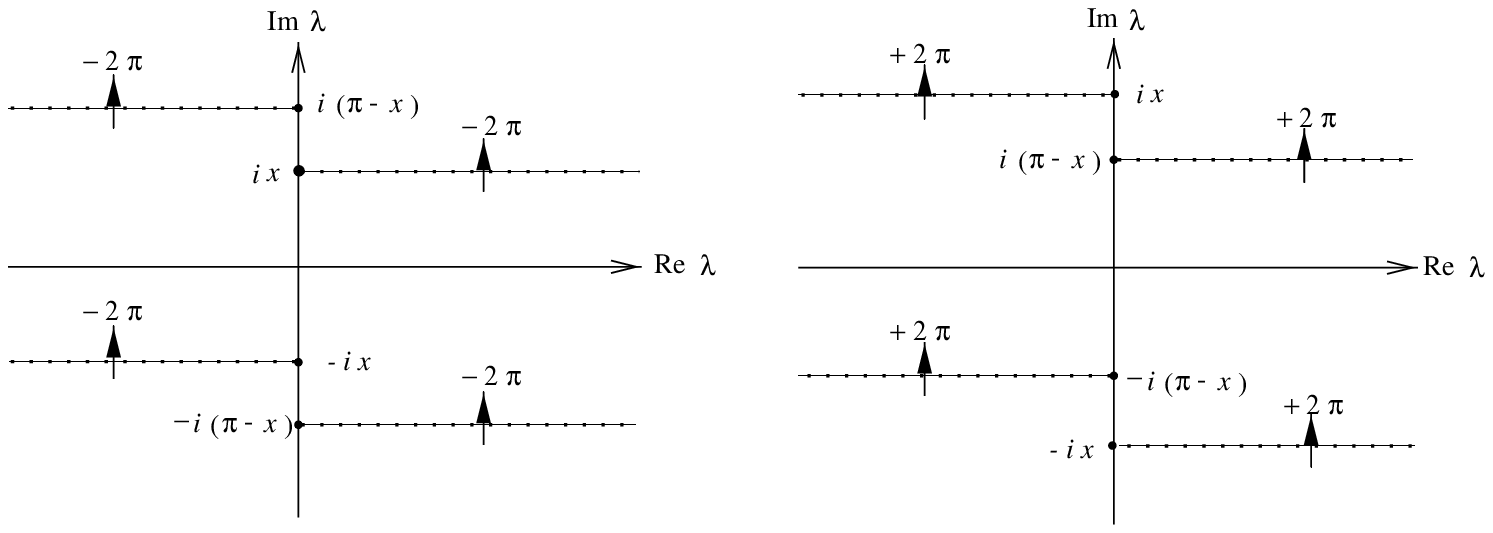,height=5.7cm} \\
\vskip .5 true cm
\noindent
Fig.1: Cuts of the function $\phi_{x/]g}(\pi\l/\g)$ 
for $x<\pi/2$ (left) and $x>\pi/2$ (right). 
\end{center}
\vskip 1 true cm
\begin{center}
       \epsfig{file=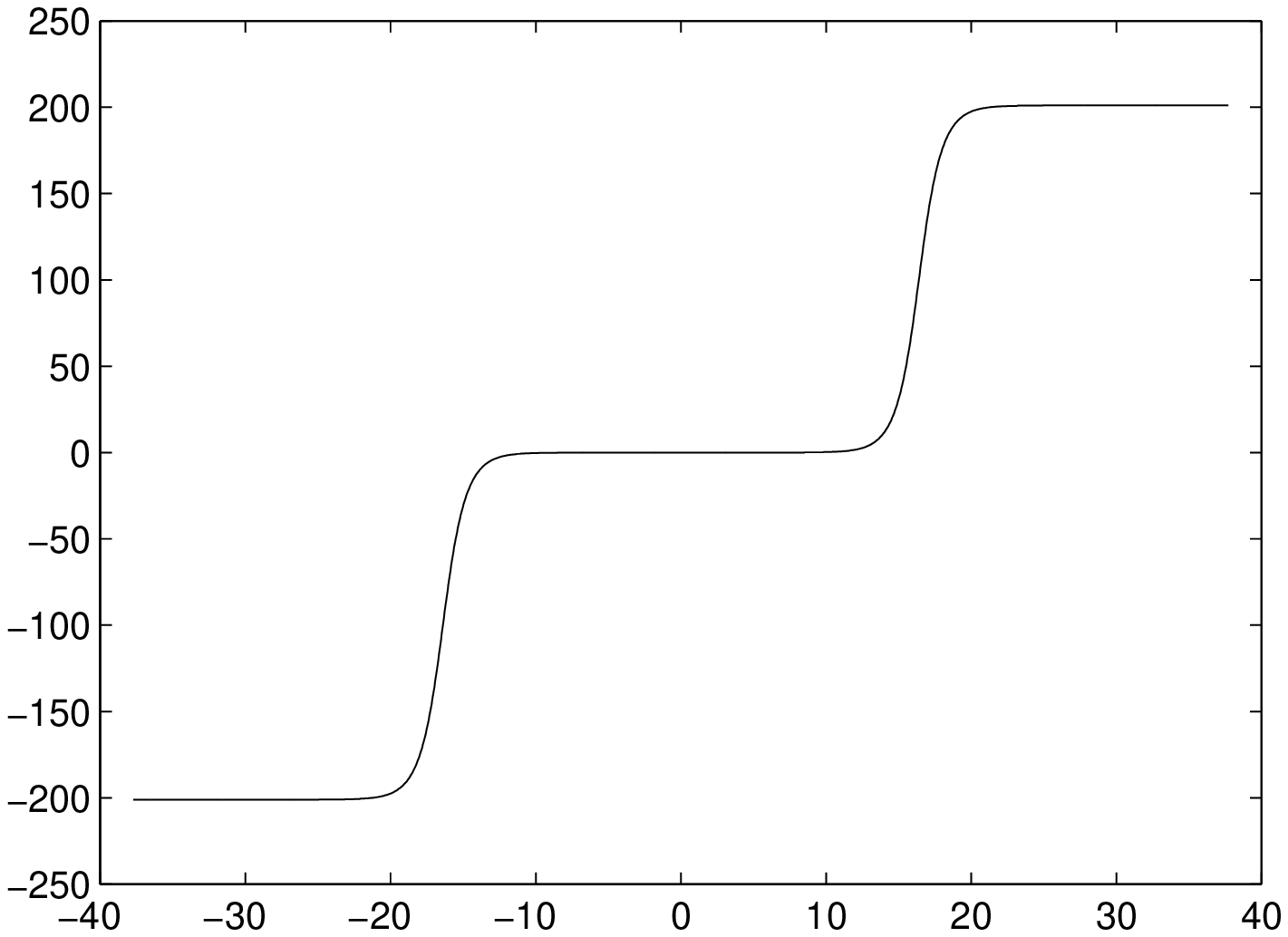,height=8cm} \\
\vskip .5 true cm
\noindent
Fig.2: The ground state counting function $Z_N(\l)$ for $N=64$,
$\g=7\pi/17$ and $\Th=16.4$. It is calculated by solving numerically
the corresponding BAE.
\end{center}
\vskip 2 true cm
\eject

\begin{center}
       \epsfig{file=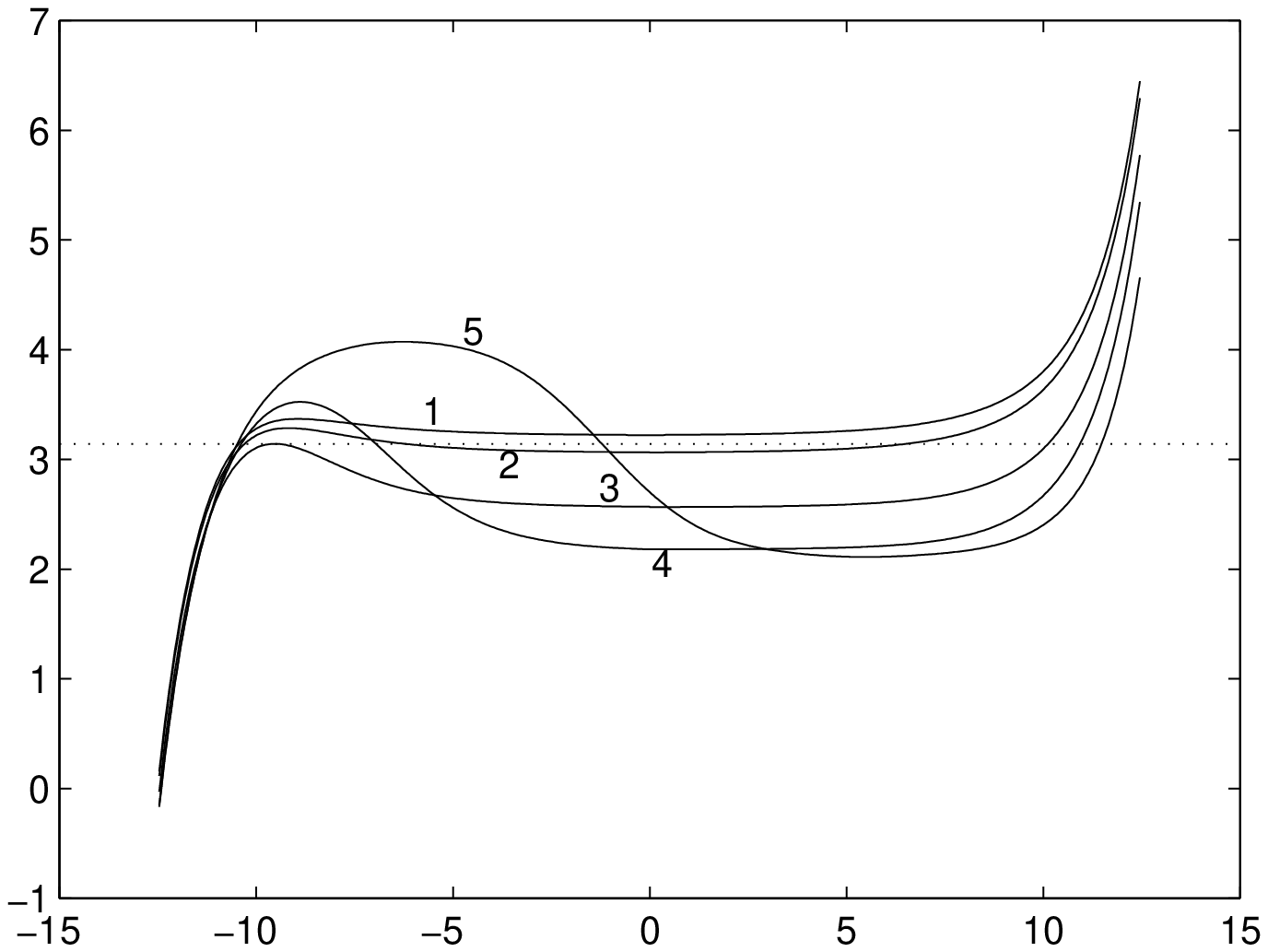,height=8cm} \\
\vskip .3 true cm
\noindent
Fig.3: The central portion of $Z_N(\l)$ for $N=64$, $\Th=16.4$,
$S^-=2$ and $S^0=S^+=0$ as $\g/\pi$ changes from 1: $\frac{10}{41}$,
2: $\frac{39}{41}$, 3: $0.296$ 4: $\frac{20}{61}$ and
5: $0.3332$. The rightmost quantization value of the
left sea is $\pi$ (the dotted line) and is occupied by a real root
which gets closer and closer to the origin. For a detailed description
see section \ref{largeTh}.
\end{center}
\vskip 2 true cm
\begin{center}
       \epsfig{file=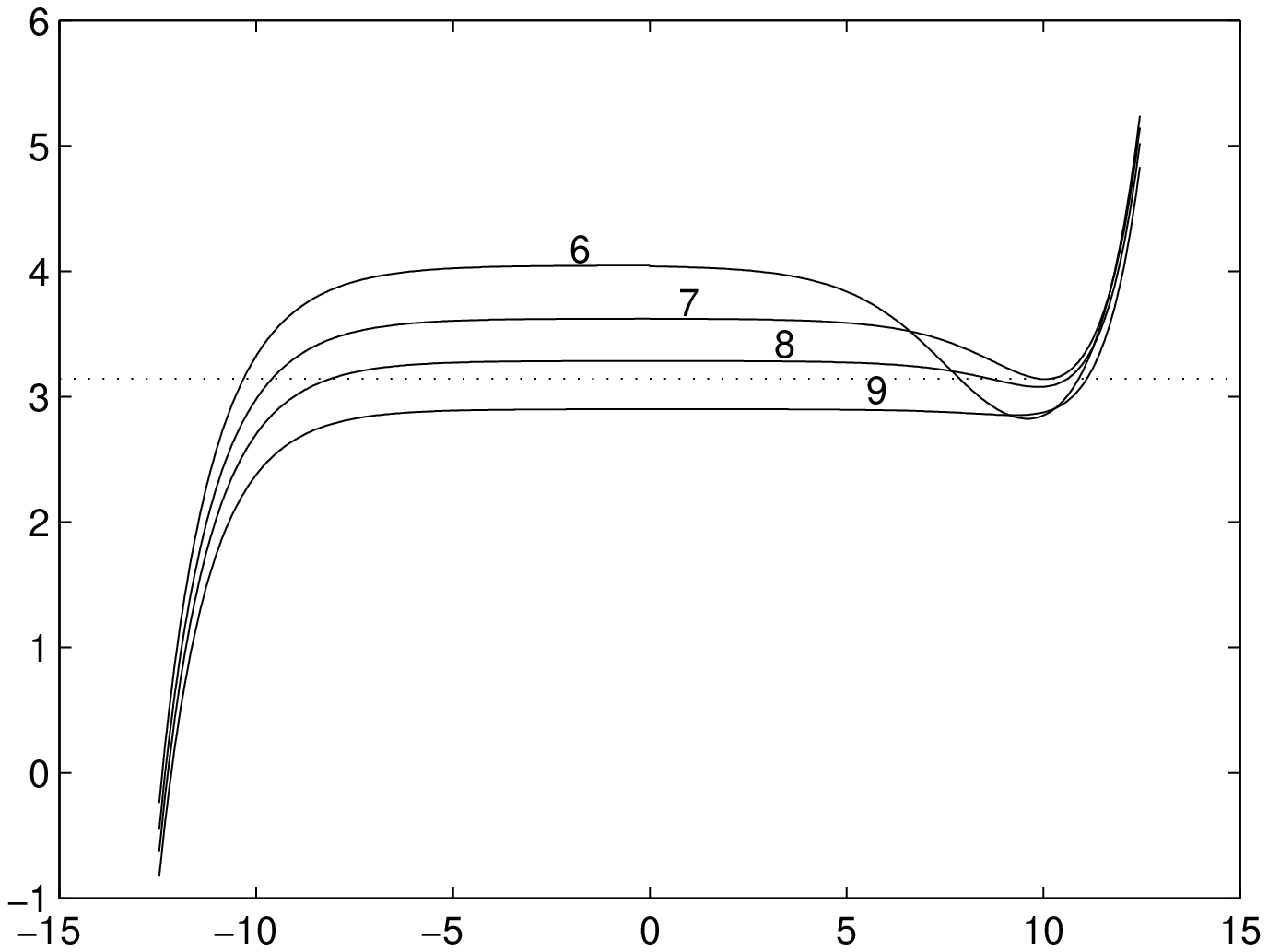,height=8cm} \\
\vskip .3 true cm
\noindent
Fig.4: The central portion of $Z_N(\l)$ for $N=64$, $\Th=16.4$,
$S^-=3$, $S^0=0$ and $S^+=-1$ as $\g/\pi$ changes from 6:
$\frac{20}{59}$, 7: $0.356$, 8: $0.369$ and 9:
$\frac{5}{13}$. The real root at $Z_N(\l)=\pi$ has passed to the right
sea. For a detailed description see section \ref{largeTh}.
\end{center}

\end{document}